\documentclass[conference]{IEEEtran}

\usepackage{color}
\usepackage[english]{babel}
\usepackage{booktabs}
\usepackage{paralist}
\usepackage{amsmath}
\usepackage{graphicx}
\usepackage{url}
\usepackage{epstopdf}
\usepackage{float}
\usepackage{cite}
\usepackage{enumerate}
\usepackage{fixltx2e}
\usepackage{algorithm}
\usepackage[noend]{algpseudocode}
\usepackage{relsize}

\usepackage[labelformat=simple, subrefformat=simple]{subcaption}

\title{QAware: A Cross-Layer Approach to MPTCP Scheduling}
\author{
\IEEEauthorblockN{Tanya Shreedhar\IEEEauthorrefmark{1}, Nitinder Mohan\IEEEauthorrefmark{2}, Sanjit K. Kaul\IEEEauthorrefmark{1}, Jussi Kangasharju\IEEEauthorrefmark{2}}
\IEEEauthorblockA{\IEEEauthorrefmark{1}Wireless Systems Lab, IIIT-Delhi, India, 
\IEEEauthorrefmark{2}University of Helsinki, Finland}
}

\makeatletter
\def\therule{\makebox[\algorithmicindent][l]{\hspace*{.5em}\vrule height .75\baselineskip depth .25\baselineskip}}%

\newtoks\therules
\therules={}
\def\appendto#1#2{\expandafter#1\expandafter{\the#1#2}}
\def\gobblefirst#1{
  #1\expandafter\expandafter\expandafter{\expandafter\@gobble\the#1}}%
\def\LState{\State\unskip\the\therules}
\def\pushindent{\appendto\therules\therule}%
\def\popindent{\gobblefirst\therules}%
\def\printindent{\unskip\the\therules}%
\def\printandpush{\printindent\pushindent}%
\def\popandprint{\popindent\printindent}%

\algdef{SE}[WHILE]{While}{EndWhile}[1]
  {\printandpush\algorithmicwhile\ #1\ \algorithmicdo}
  {\popandprint\algorithmicend\ \algorithmicwhile}%
\algdef{SE}[FOR]{For}{EndFor}[1]
  {\printandpush\algorithmicfor\ #1\ \algorithmicdo}
  {\popandprint\algorithmicend\ \algorithmicfor}%
\algdef{S}[FOR]{ForAll}[1]
  {\printindent\algorithmicforall\ #1\ \algorithmicdo}%
\algdef{SE}[LOOP]{Loop}{EndLoop}
  {\printandpush\algorithmicloop}
  {\popandprint\algorithmicend\ \algorithmicloop}%
\algdef{SE}[REPEAT]{Repeat}{Until}
  {\printandpush\algorithmicrepeat}[1]
  {\popandprint\algorithmicuntil\ #1}%
\algdef{SE}[IF]{If}{EndIf}[1]
  {\printandpush\algorithmicif\ #1\ \algorithmicthen}
  {\popandprint\algorithmicend\ \algorithmicif}%
\algdef{C}[IF]{IF}{ElsIf}[1]
  {\popandprint\pushindent\algorithmicelse\ \algorithmicif\ #1\ \algorithmicthen}%
\algdef{Ce}[ELSE]{IF}{Else}{EndIf}
  {\popandprint\pushindent\algorithmicelse}%
\algdef{SE}[PROCEDURE]{Procedure}{EndProcedure}[2]
   {\printandpush\algorithmicprocedure\ \textproc{#1}\ifthenelse{\equal{#2}{}}{}{(#2)}}%
   {\popandprint\algorithmicend\ \algorithmicprocedure}%
\algdef{SE}[FUNCTION]{Function}{EndFunction}[2]
   {\printandpush\algorithmicfunction\ \textproc{#1}\ifthenelse{\equal{#2}{}}{}{(#2)}}%
   {\popandprint\algorithmicend\ \algorithmicfunction}%
\makeatother

\algnewcommand{\Inputs}[1]{%
  \State \textbf{Inputs:}
  \Statex \hspace*{\algorithmicindent}\parbox[t]{.8\linewidth}{\raggedright #1}
}
\algnewcommand{\Initialize}[1]{%
  \State \textbf{Initialize at packet arrival $P_k$:}
  \Statex \hspace*{\algorithmicindent}\parbox[t]{.8\linewidth}{\raggedright #1}
}

\IEEEoverridecommandlockouts

\IEEEpubid{%
   \makebox[\columnwidth]{ISBN 978-3-903176-08-9~\copyright~2018 IFIP \hfill}%
   \hspace{\columnsep}%
   \makebox[\columnwidth]{ }%
}

\begin{document}

\maketitle

  \begin{abstract}

Multipath TCP (MPTCP) allows applications to transparently use all available network interfaces by creating a TCP subflow per interface. One critical component of MPTCP is the scheduler that decides which subflow to use for each packet. Existing schedulers typically use estimates of end-to-end path properties, such as delay and bandwidth, for making the scheduling decisions. In this paper, we show that these scheduling decisions can be significantly improved by including readily available local information from the device driver queues to the decision-making process. We propose QAware, a novel cross-layer approach for MPTCP scheduling. QAware combines end-to-end delay estimates with local queue buffer occupancy information and allows for a better and faster adaptation to the network conditions. This results in more efficient use of the available resources and considerable gains in aggregate throughput. We present the design of QAware and evaluate its performance through simulations, and also through real experiments, comparing it to existing schedulers. Our results show that QAware performs significantly better than other available approaches for all use-cases and applications.

\end{abstract}  


\section{Introduction}\label{sec:introduction}
 
Multipath TCP (MPTCP) is a recently-standardized extension to TCP that allows devices with multiple network interfaces, e.g., smartphones with WiFi and LTE, to seamlessly form multiple parallel connections to exploit the full network capacity. MPTCP offers increased robustness and resilience, as well as seamless handovers and it has been proposed to be also used in datacenters~\cite{datacentermptcp}, opportunistic networks~\cite{opportunisticmptcp}, etc. There is both an open source implementation for Linux~\cite{mptcplinux}, and companies, such as Apple, have incorporated MPTCP into their products and have made the APIs open to application developers~\cite{applemptcp}.

\begin{figure}[!t]             
\centering
\includegraphics[width=0.35\textwidth]{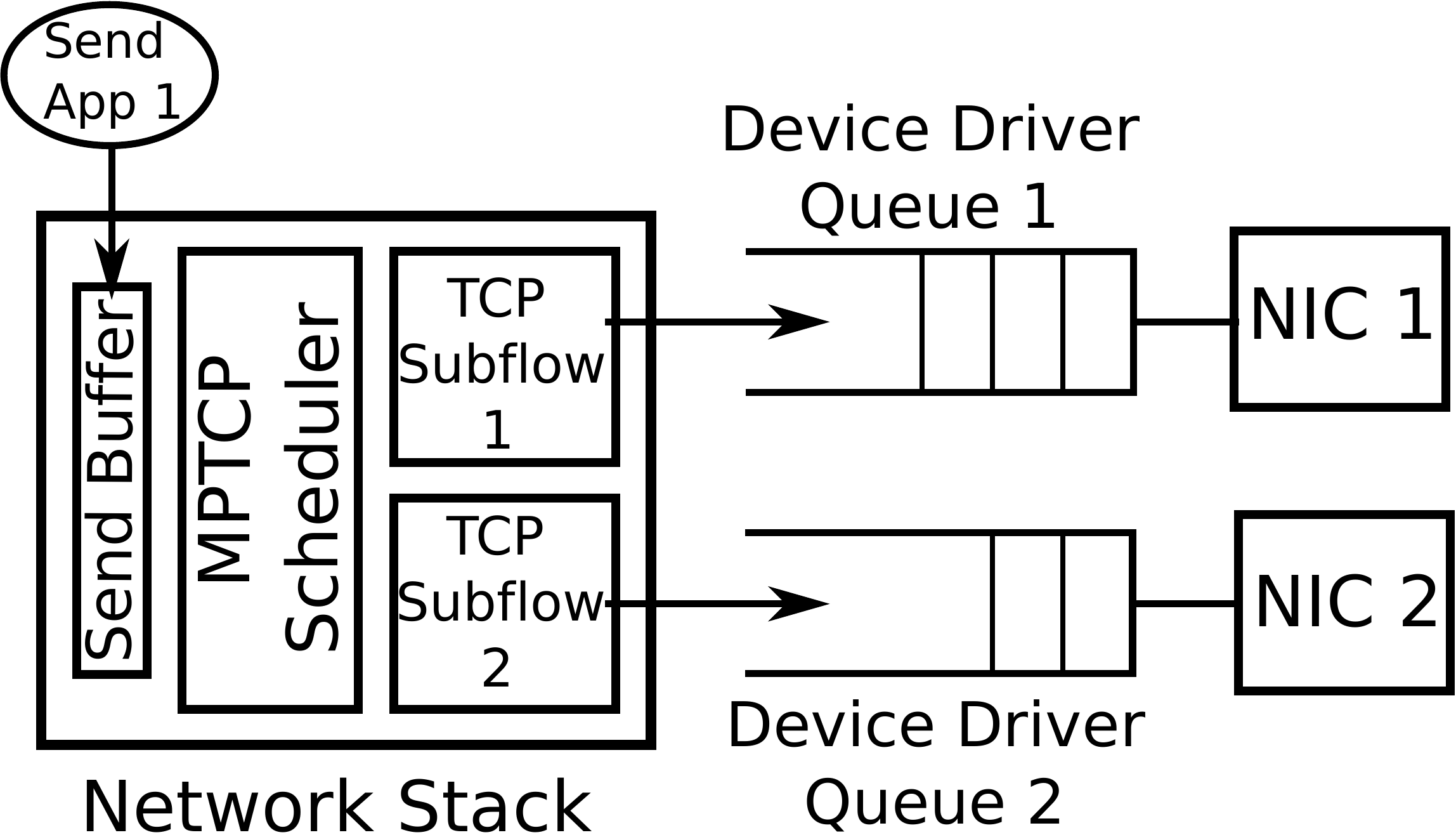}
\caption{\label{fig:mptcp_device_queue} An illustration of MPTCP-compliant machine and how its subflows interact with their corresponding network interface queues.}
\end{figure}

Figure~\ref{fig:mptcp_device_queue} shows the network stack of MPTCP-compliant machine. Applications utilizing MPTCP can send their data over multiple TCP subflows, where each subflow is associated with a unique network interface. TCP packets scheduled over a subflow wait in the device driver queue of the corresponding network interface before they are transmitted by the network interface card (NIC). The choice of network path for sending application data is made by the MPTCP scheduler block and depends on the scheduling policy.

Scheduling between the multiple connections is an obvious research problem and recently multiple proposals~\cite{blest},~\cite{dems},~\cite{daps},~\cite{ecf} have emerged to improve the  default MPTCP scheduler~\cite{mptcpscheduler}. Typically, these schedulers use a transport layer estimate of the end-to-end bandwidth/delay (for example, the smoothed round-trip time) for each TCP subflow as an input to the scheduling policy that decides on how application data must be assigned to the multiple subflows.

In this paper, we propose a novel scheduler for MPTCP, QAware, which departs from the previous proposals in a fundamental way. While we also use the end-to-end delay estimates, like current schedulers, QAware additionally considers the number of packets in the device driver queue of the sender. This modification is motivated by our findings, which we discuss further in Section~\ref{sec:incognizance}. The key motivation, as we will demonstrate, is that as a particular flow is used more, its end-to-end delay increases gradually, making it less attractive to use. However, the traditional, purely end-to-end-based estimation, reacts very slowly to these changes.

Additionally, utilizing queue occupancy information allows QAware to use all available subflows optimally, especially when their properties are highly heterogeneous. Existing proposals like~\cite{blest, ecf, otias}, treat the flows as separate entities and typically do not fully use all the flows. 
QAware optimizes transmission over all the flows and gets a significantly higher aggregate throughput, with no loss of performance in any situation.

The contributions of this paper are:

\begin{enumerate}[(a)]
\item We propose QAware, which is a novel cross-layer approach to scheduling packets across all available MPTCP subflows. The design is motivated by our experimental findings that combining local device driver queue occupancy with the traditional end-to-end delay measurements yield far superior performance.

\item We model available MPTCP subflows as multiple parallel service facilities that can service data provided by an application. This enables us to leverage queueing theoretical insights to create a scheduling policy that combines end-to-end delays and device driver queue occupancy.
\item Our simulations and real-world experimentation over a wide range of applications compare QAware with the default MPTCP scheduler~\cite{mptcpscheduler}, ECF~\cite{ecf}, DAPS~\cite{daps}, and BLEST~\cite{blest}.

\end{enumerate}

Rest of the paper is organized as follows. We discuss the relevant background and related works in Section~\ref{sec:relatedwork}. Section~\ref{sec:incognizance} motivates the need for a cross-layer approach to scheduling. In Section~\ref{sec:scheduler}, we describe the scheduling policy used by QAware. Section~\ref{sec:implementation} provides implementation details of QAware in latest MPTCP v0.93. Section~\ref{sec:evaluation} provides an overview of our evaluation methodology. Sections~\ref{sec:simulation} and~\ref{sec:realexperiments} quantify the performance of QAware using extensive simulations and real-world experiments, respectively. We conclude in Section~\ref{sec:conclusion}.

\section{Background and Related Work}\label{sec:relatedwork}

\begin{figure}[!t]
\centering
\begin{subfigure}{.23\textwidth}
\centering
  \includegraphics[width=\linewidth]{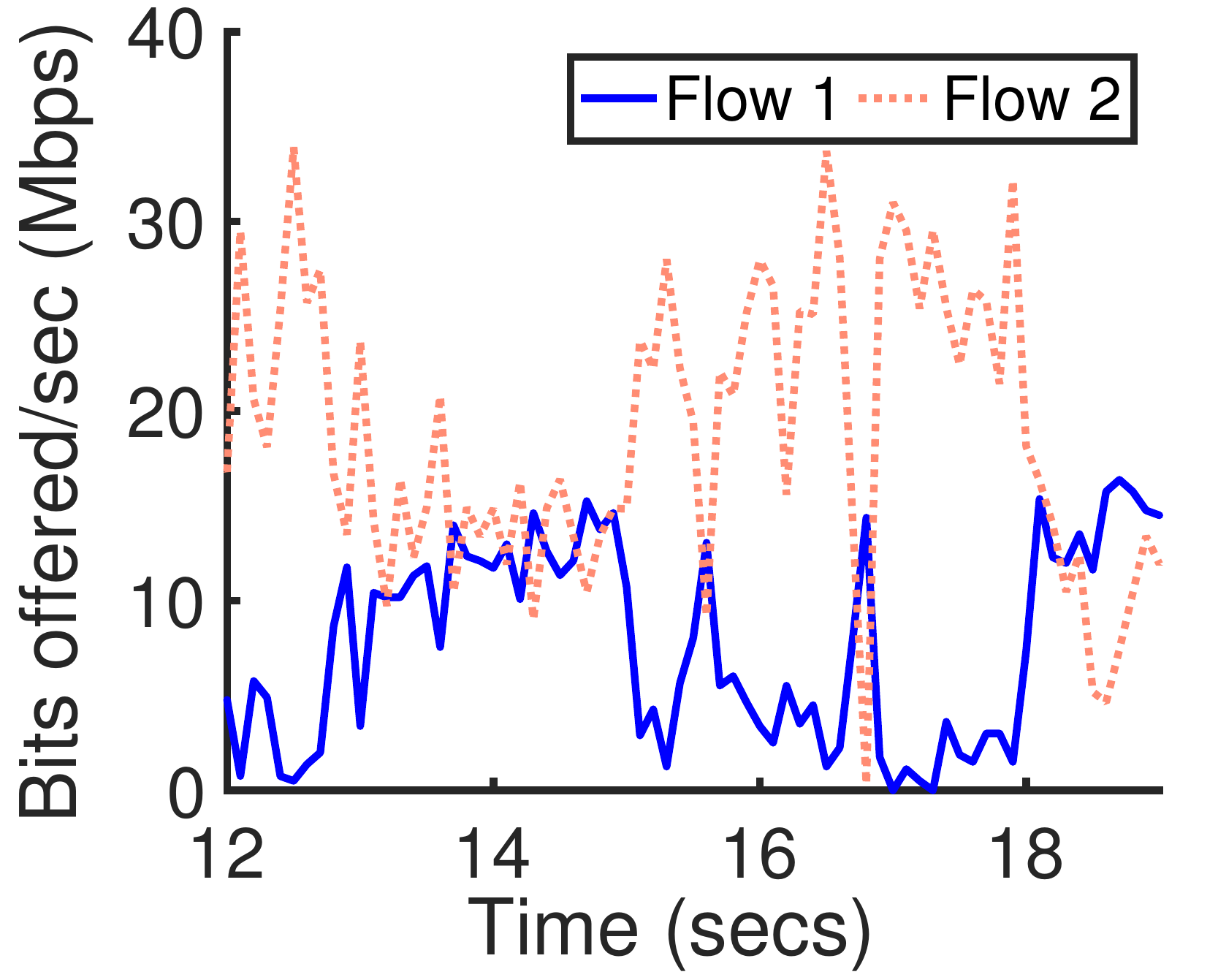} 
  \caption{\label{fig:emc_throughput}}
\end{subfigure}%
\hspace{0.005\textwidth}
\begin{subfigure}{.22\textwidth}
\centering
  \includegraphics[width=\linewidth]{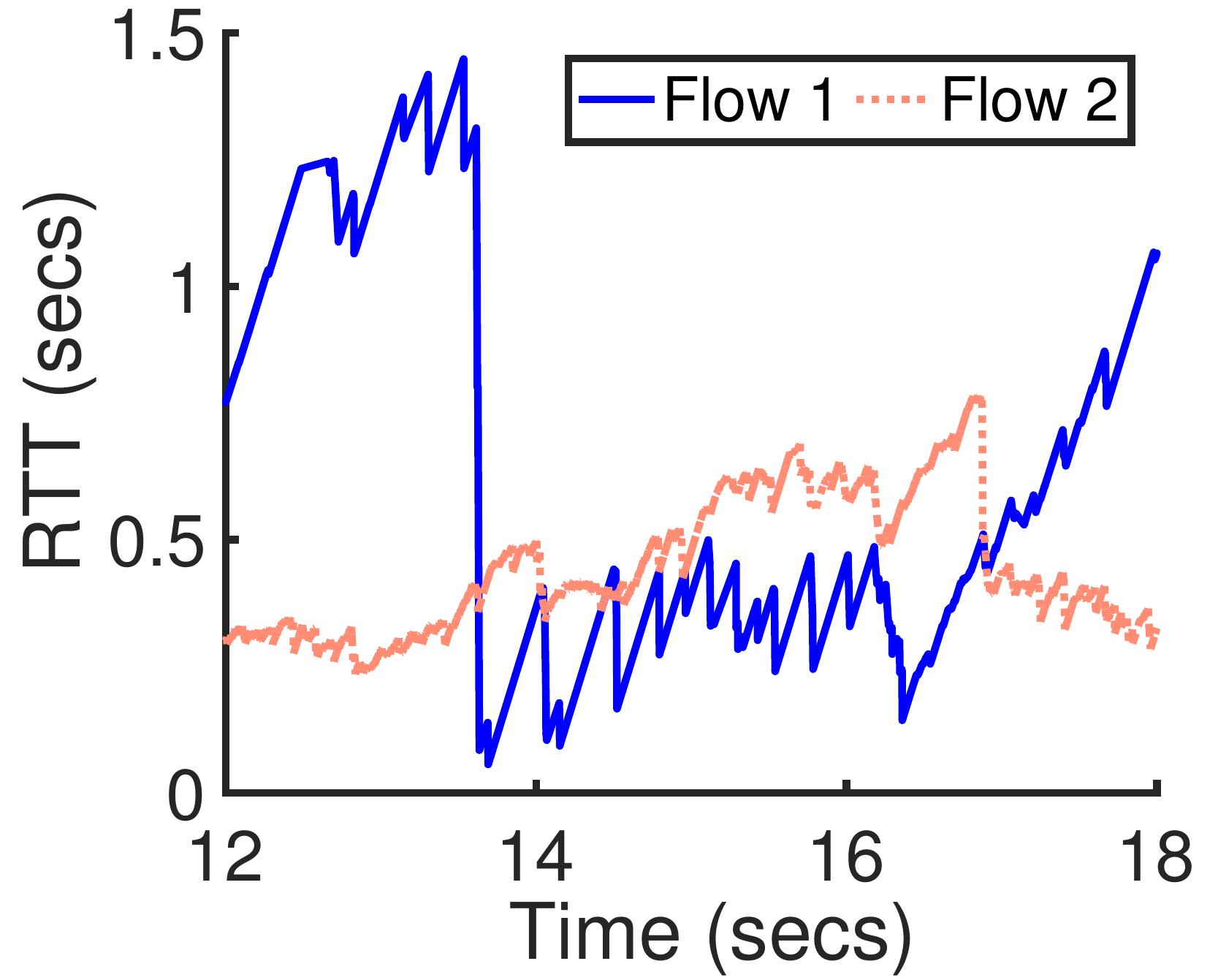} 
  \caption{\label{fig:emc_rtt}}
\end{subfigure}
\caption{(a) Loading (Mbps) at the subflows and (b) their RTT(s). The paths taken by the subflows and the network are shown in Figure~\ref{fig:netw_topology}.}
\label{fig:emc_results}
\end{figure}

The default MPTCP scheduler (minSRTT) allocates traffic on the fastest subflow (one that has the smallest smoothed RTT) with available congestion window at each packet arrival. Several researchers have proposed improvements to the default minSRTT scheduler. Most approaches leverage the difference in RTT of the subflows \cite{slow-path-adaptation, packet-scheduling}. Others have also considered additional TCP-layer parameters such as SSThresh, congestion window, selective ACK and receiver buffer size along with RTT \cite{receiver-scheduler, ocps, f2dpds}. 

In \cite{otias}, the authors introduce an additional sender queue to schedule packets on a subflow even when it is unavailable. Delay Aware Packet Scheduler (DAPS) \cite{daps} generates a schedule for sending future segments over subflows based on their RTT ratios. However, this makes DAPS unable to react promptly to network changes due to pre-computed long schedules.
Blocking-Estimation-based MPTCP Scheduler (BLEST) \cite{blest} aims to reduce head-of-line blocking by waiting for the faster subflow despite the availability of space in congestion window of the slower subflow. ECF \cite{ecf} follows a similar principle as that of BLEST, but while BLEST aims to reduce out-of-order delivery assuming that the send buffer is a bottleneck, ECF aims to minimize completion time.


Researchers have also proposed schedulers that improve MPTCP performance for specific application use-cases. Decoupled Multipath Scheduler (DEMS) \cite{dems} aims to reduce fixed-size file's delivery time over MPTCP by estimating available bandwidth on subflows. However, the authors rely on exact knowledge of data chunk boundary for efficient scheduling.
In \cite{crosslayer_video} authors leverage application layer information for flow scheduling decisions to provide delay-resilient video streaming in MPTCP. MP-DASH \cite{mp-dash} exploits path information from streaming client to improve DASH video delivery.
\cite{crosslayer_infocomm} labels WiFi subflow as active/inactive for data transmission based on a minimum desired signal strength. However, unlike other cross-layer approaches which optimize specific application performance over MPTCP, QAware taps into lower layer information to improve performance for \emph{all} MPTCP traffic. Furthermore, as shown later in the paper, QAware's unique design of leveraging hardware queue occupancy enables it to swiftly adapt to varying network conditions and co-existent network applications sharing bottleneck paths. 

\section{Motivating Use of Cross-Layer Information} \label{sec:incognizance}

\begin{figure}[!t]             
\centering
\includegraphics[width=0.4\textwidth]{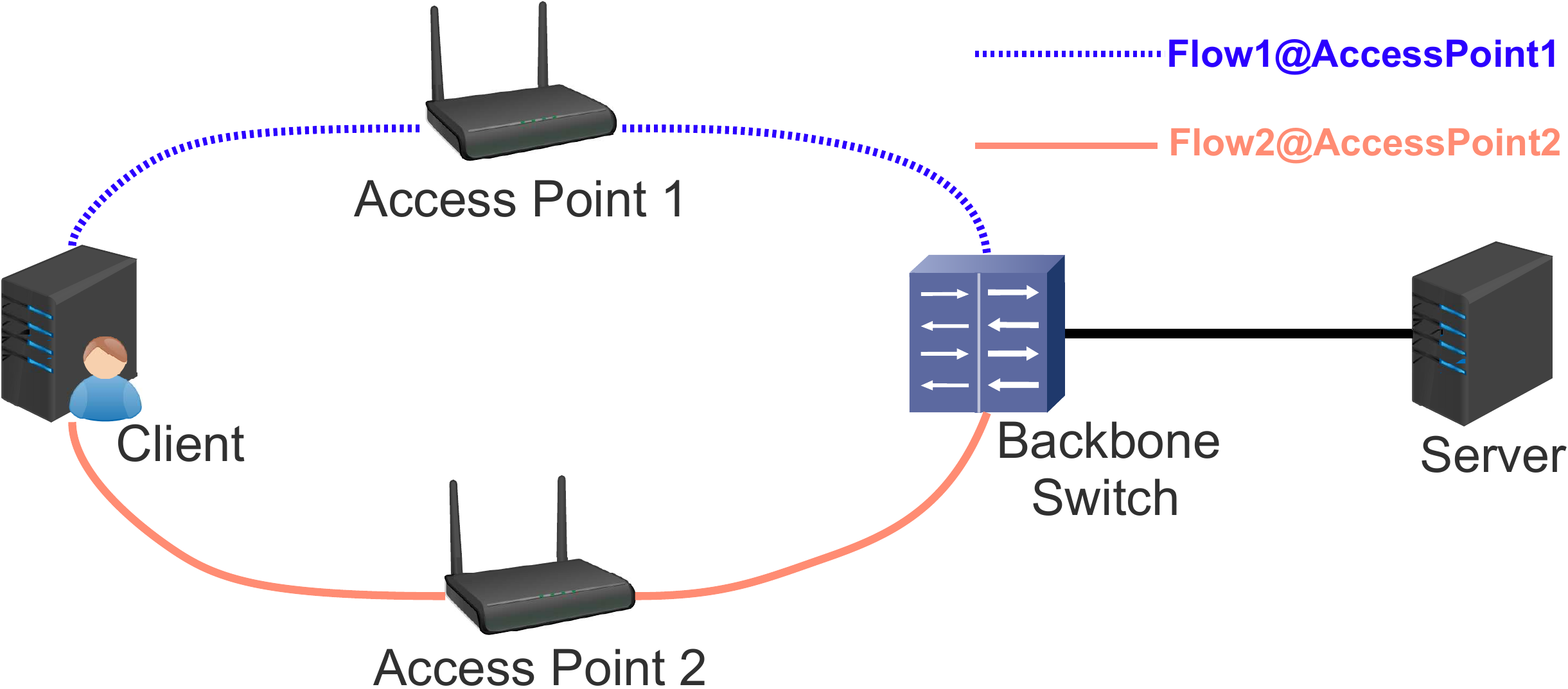}
\caption{Topology used in experiments and simulations.}
\label{fig:netw_topology} 
\end{figure}

Figures~\ref{fig:emc_throughput} and~\ref{fig:emc_rtt} respectively show loading (bits offered per second) and the corresponding estimates of round-trip times (RTT) of two available subflows by the default MPTCP scheduler, minSRTT. They were obtained from controlled testbed experiments and show how the scheduler optimizes over two available TCP subflows that use non-interfering end-to-end paths. The network topology used in the experiment is shown in Figure~\ref{fig:netw_topology}. The last-mile links were WiFi using 802.11g and the rest were $1$ Gbps Ethernet. Neither flow dropped any packets during the length of the experiment. 

In the experiment, the default scheduler only utilizes $\approx 60\%$ of available aggregated bandwidth. Observe (Figure~\ref{fig:emc_throughput}) that the default scheduler, more often than not, prefers to send packets on one flow over the other. However, this by itself is not responsible for the low utilization of the available bandwidth. The reason, we argue, is that the default scheduler loads a flow deemed to be the best amongst available flows for \textit{undesirably long} intervals. This is because the scheduler uses only the SRTT of the flows, which is a delayed end-to-end transport layer measurement, for its scheduling decisions. 

Consider the RTT of flow $1$ in Figure~\ref{fig:emc_rtt}. The RTT captures in a lagged manner the impact of scheduling decision on the subflow. The consistently high values (see interval $12$s to $14$s in the figure) correspond to an earlier interval of time when the subflow was being assigned packets by the scheduler while it was heavily loaded. That is, the device queue corresponding to the subflow had previously many packets queued at the NIC. 

The sharp dip in values (around time $14$s in the figure) captures the transition from when the flow stopped being assigned packets due to high RTT to when it was again assigned packets. These assigned packets arrive at a rather lightly loaded flow and see much smaller RTT, which causes the dip. The small RTT that follows the dip corresponds to packets being assigned to the flow while it was still lightly loaded. As the subflow continues to be assigned packets, the same is reflected, albeit in a delayed manner, in increasing RTT (seconds $16$ to $18$ in Figure~\ref{fig:emc_rtt}) that eventually peaks as it did during $12-14$ seconds. By the time the resulting large RTT makes the scheduler switch to the other flow, the scheduler has already spent an undesirably long time injecting packets to a loaded subflow.

In summary, the scheduling decisions that lead to high device queue occupancy and increase in RTT were made using values of RTT that corresponded to an earlier interval when the flow was less loaded. So while a device queue (local to the MPTCP sender and used by the MPTCP flow) is loaded with packets, MPTCP scheduler remains oblivious to the same. Instead, it waits to be informed via a delayed end-to-end RTT based feedback mechanism. In the process, it loses out on many opportunities of scheduling packets to the other better flow; one that is lightly loaded.

The above observations motivate QAware. It uses the occupancy of the device queues together with RTT estimates to use all available flows more efficiently.

\section{QAware Scheduler} \label{sec:scheduler}
\begin{figure}[!t]             
\begin{center}
\includegraphics[width=0.4\textwidth]{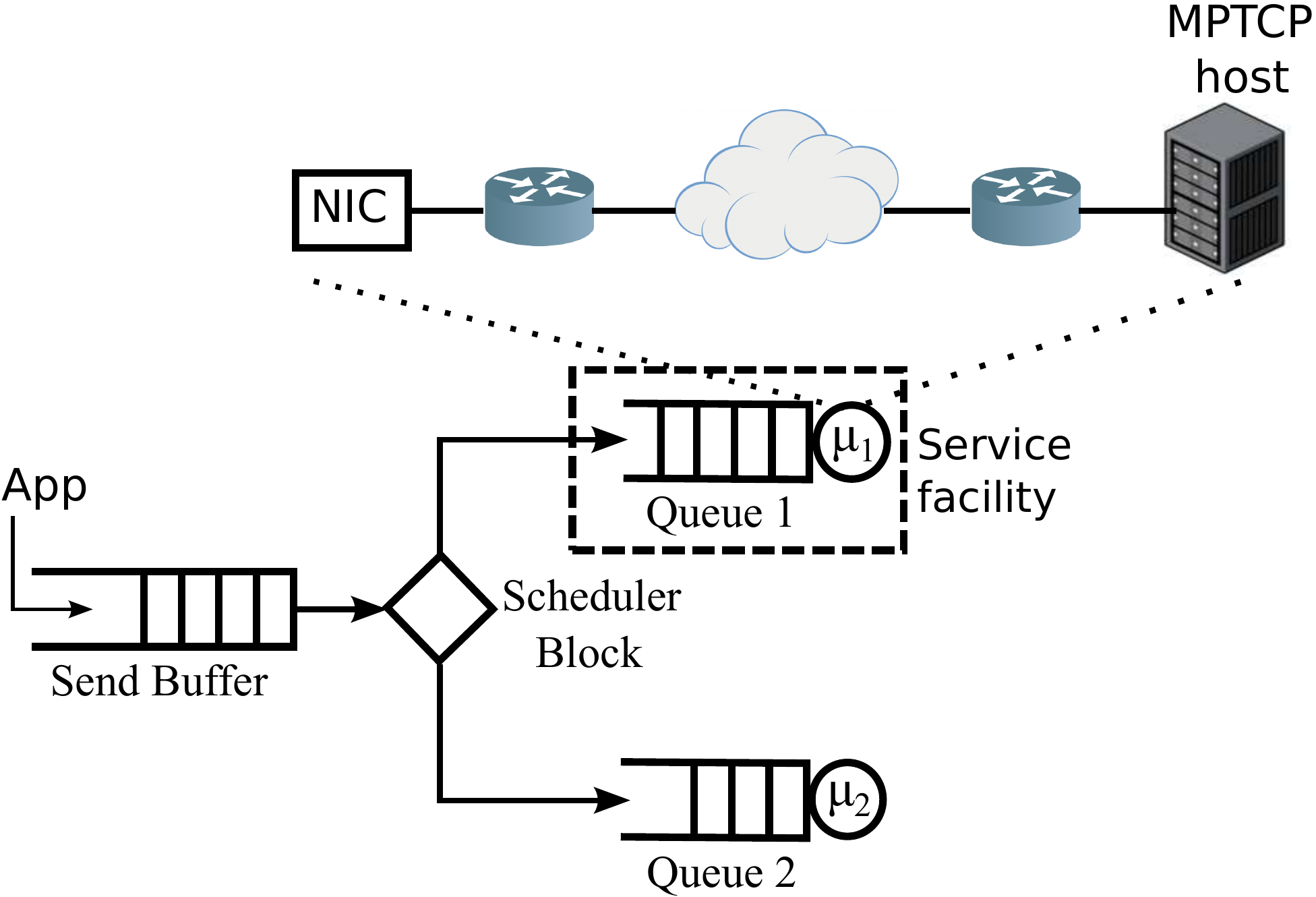}
\caption{Queueing abstraction of an \emph{end-to-end} MPTCP connection with two subflows.}
\label{fig:networkqueue}
\end{center}
\end{figure}
We consider a simplified queue-theoretic abstraction to capture the essentials of the scheduling problem, with the goal of maximizing \emph{end-to-end} throughput. Specifically, we model each subflow by a service facility. Figure~\ref{fig:networkqueue} illustrates the abstraction for an MPTCP \emph{end-to-end} connection that uses two TCP flows. The abstraction allows us to apply results from analysis of multi-queue systems\cite{rosberg}.

In our queueing abstraction, packets generated by an application arrive into a queue that models the TCP send buffer (Figure~\ref{fig:mptcp_device_queue}). Packets in this queue are assigned to one of the available service facilities in a first-come-first-serve (FCFS) manner. Each facility consists of a finite queue and a server. Packets inside a facility are serviced in an FCFS manner. 

The queue in a service facility is the device driver queue (Figure~\ref{fig:mptcp_device_queue}) that is used by the TCP subflow corresponding to the facility. The server includes the source host NIC, access network used by the subflow, intermittent nodes in the core and the destination host (all layers of the TCP/IP stack).

When a packet is assigned to a service facility, it may find other packets waiting for service in the facility's queue. This packet must wait for all the other waiting packets to finish service before it enters the server of the facility. The total time a packet spends in a facility, often referred to as its \emph{system time}, includes the time it waits in the facility's queue and the time it spends getting service.

\emph{Origins of the QAware scheduler:} Many analytical works on queueing systems have looked at scheduling customer/packet arrivals to parallel service facilities\cite{rosberg,weber,whitt,winston}. For many general arrival processes and service time distributions, when all servers are stochastically identical, the optimal policy is to choose a service facility with a minimum number of packets in its queue~\cite{rosberg,weber,winston}, that is it minimizes the average packet system time. 
For the case of non-identical servers, a scheduling policy that assigns a packet to a service facility that minimizes the conditional expected system time of the packet, conditioned on the knowledge of the number of packets waiting for service in the facility, shows good performance~\cite{rosberg}. Our QAware scheduler uses the policy in an MPTCP setting.

Consider $K$ service facilities indexed $1,\ldots,K$. Let facility $k$ have a service rate of $\mu_k$. The two facilities in Figure~\ref{fig:networkqueue} have service rates of $\mu_1$ and $\mu_2$. Let $n_k(t)$ be the number of packets waiting for service in facility $k$ at time $t$. The policy assigns a packet to a service facility $k^*$ given by
\begin{align}
k^* = \arg \min_k \frac{n_k(t)+1}{\mu_k}.
\label{eqn:policy1}
\end{align}
Note that $1/\mu_k$ is the expected service time of a packet in facility $k$. Thus, the conditional waiting time of a packet that enters such a facility is ${n_k(t)}/{\mu_k}$, which is the sum of the expected service times of the $n_k(t)$ packets currently waiting for service in the facility. In addition, we add the term $1/\mu_k$ to ${n_k(t)}/{\mu_k}$, to include the expected service time of the packet to be scheduled. Thus, the expression being minimized in~(\ref{eqn:policy1}) is the conditional expected \emph{system time} of a packet if it were to be assigned to facility $k$.

\emph{Adapting scheduling policy~(\ref{eqn:policy1}) to multiple end-to-end TCP subflows:} The number $n_k(t)$ of packets in the queue of service facility $k$ is the number of packets waiting in the device driver queue of the corresponding subflow $k$ and can be obtained. However, we must estimate the average service time $1/\mu_k$ of subflow $k$.

Consider the $i$\textsuperscript{th} packet arrival. Let $t_i^s$ be the time the packet is assigned to a subflow. Let $t_i^a$ be the time that a TCP ACK acknowledges receipt of the packet. The round-trip time of the packet is $\text{RTT}_i = t_i^a - t_i^s$. Note that this includes the time packet waits in the device driver queue of its assigned subflow before it starts service and the time it spends in service. This is the \emph{system time} of the packet. Let $W_i$\footnote{For simplicity of exposition we ignore the time a TCP ACK may have to wait in a queue before being sent to the TCP layer.} be the time the packet $i$ waits in the queue. This time can be calculated locally at the MPTCP sender. The time $X_i$ that the packet spends in service begins when the packet enters the NIC for transmission and ends when a TCP ACK for the packet is received. Given $W_i$ and $\text{RTT}_i$, we have $X_i = \text{RTT}_i - W_i$. The estimate of the service time is updated on receipt of a TCP ACK. Let $\hat{S}_k$ be the current estimate of the average service time of facility $k$. On receipt of a TCP ACK for packet $i$, we update
\begin{align}
\hat{S}_k = \alpha \hat{S}_k + (1-\alpha) X_i,
\label{eqn:hatSk}
\end{align}
where $0 < \alpha < 1$ applies appropriate weights to the last estimate of the average and the current service time. We use $\alpha=0.8$ in this work which is also the smoothing factor for TCP congestion control \footnote{We examined for other values of $\alpha$ which did not impact the overall performance of QAware.} .
The corresponding estimate of the service rate is $1/\hat{S}_k$. At time $t$, QAware schedules to the TCP subflow $k^*$ that satisfies
\begin{align}
k^* = \arg \min_k {(n_k(t) + 1)} \hat{S}_k.
\label{eqn:policy3}
\end{align}
Finally, note that since $X_i = \text{RTT}_i - W_i$, we have $\hat{S}_k = \text{RTT} - \widehat{W}$, where RTT and $\widehat{W}$ are the exponentially weighted moving averages, with coefficient $\alpha$, of packet round-trip times and device driver queue waiting times, respectively, for the subflow $k$. In our real implementation, summarized in (Algorithm~\ref{alg:queueaware}), we use RTT estimates that are readily available for each subflow and we calculate an approximation of $\widehat{W}$ based on information available from device driver queues.
\begin{algorithm}[!t]\small
\caption{QAware Algorithm}\label{alg:queueaware}
\begin{algorithmic}[1]
\Inputs{Available Subflows \texttt{SF$\in \{1,\ldots,n\}$}}
\Initialize{\strut\texttt{minService} $\gets$ \texttt{0xFFFFFFFF} \\ 
\texttt{selectedSubflow} $\gets$ NONE}
\item[]
\LState //The function below will return best subflow for packet $P_k$
\For{each \texttt{$\text{subflow} \in$SF}}
	\LState $n_k \gets$ \texttt{queueSize(subflow)}
	\If{\texttt{$n_k \not= 0$}}
		\LState $\Delta t$ $\gets$ sampling time
		\LState $\Delta $packets $\gets$ packets dequeued in $\Delta t$
		\LState $W_k \gets [1/(\frac{\text{$\Delta $packets}}{\text{$\Delta t$}})] n_k$
	\Else
		\LState $W_k \gets 0$
	\EndIf
	\LState $\widehat{W} \gets \alpha \widehat{W} + (1-\alpha) W_k$
	\LState $\hat{S}_k = [\text{RTT} -  \widehat{W}]$
	\LState $TS_k = (n_k + 1) \hat{S}_k$
	\If {\texttt{$ TS_k < \text{minService}$}}
		\LState $\texttt{minService} \gets TS_k$
		\LState \texttt{selectedSubflow $\gets$ subflow}
	\EndIf
\EndFor
\end{algorithmic}
\end{algorithm}

\section{Implementation}
\label{sec:implementation}

We implement QAware as a modular scheduler using MPTCP v0.93 based on Linux kernel v4.9.60 \cite{mptcplinuximpl}. The code is available at \cite{queueawarecode}.

As shown in Section~\ref{sec:scheduler}, QAware's functioning depends on the current estimate of network interface (NIC) queue occupancy. Conventionally, the NIC queues were either implemented within the hardware itself or as part of the driver; which made NIC queues invisible to the Kernel and its occupancy extremely hard to estimate. However, since Linux Kernel $>$ v3.3.0, several NIC queue management protocols, known as Byte Queue Limits (BQL), have been introduced as part of the Kernel code to resolve starvation and latency at the NIC \cite{ietfbql}. The BQL algorithms push queueing abstractions from hardware drivers to specific data structures which can be accessed from within the Kernel \footnote{Currently, only PCIe-based ethernet drivers support BQL \cite{bqldrivers}. However, a significant effort is being made from the Linux developer community to support broader list of NICs, including wireless NIC's \cite{makewififast}.}. 

Our implementation closely follows the Algorithm \ref{alg:queueaware}. We first tap the network device address mapped to MPTCP socket via \texttt{struct dst\_entry} to access \textit{DQL}\footnote{In Linux, BQL is implemented as Dynamic Queue Limit (DQL).} as follows:
\begin{align*}
\texttt{
dql = netdev\_get\_tx\_queue(dst->dev)->dql
}
\end{align*}

We further utilize \textit{DQL} entry to estimate current NIC (\textit{netdevice}) queue occupancy of each MPTCP subflow.
\begin{equation*}
\begin{split}
\texttt{qSize} = \{\texttt{dql->num\_queued -} \\
 \hfill \texttt{dql->num\_completed}\}
\end{split}
\end{equation*}

Here, \texttt{num\_queued} and \texttt{num\_completed} refer to the total number of bytes queued in the network device and number of bytes successfully transmitted by the device respectively.

Apart from NIC queue estimates, we utilize the smoothed mean RTT estimates in microseconds via \texttt{srtt\_us} accessible through \texttt{struct tcp\_sock}. We ensure that our implementation is in line with guidelines mentioned in RFC 6182 \cite{mptcprfc}.

\section{Evaluation Methodology}\label{sec:evaluation}
In following sections, we evaluate QAware's performance through an extensive set of simulations and real-world experiments. We model our evaluation methodology to mimic real MPTCP network configurations and application use-cases. In majority of our evaluation, we model a realistic network scenario (as illustrated in Figure \ref{fig:netw_topology}) wherein a client leverages two distinct network paths to connect to a distant server. 

For simulations, we implement QAware on ns-3 network simulator. We compare QAware with default minSRTT and Earliest Completion First (ECF)~\cite{ecf} scheduler for constant bit rate (CBR), file downloads, and web browsing workloads. The simulations help us zoom into the workings of the schedulers and allow us to evaluate QAware over a variety of workloads and network path configurations. Our evaluation setup and results are described in Section \ref{sec:simulation}. 

We further examine and validate the performance gains obtained by QAware in simulated environments via real network experiments. We utilize our Kernel implementation summarized in Section \ref{sec:implementation}. The experiments were performed in a university data center and consider a variety of workloads such as video streaming, web file downloads, etc. We compare QAware with several state-of-the-art schedulers such as minSRTT, Delay Aware Packet Scheduler (DAPS)~\cite{daps}, Blocking Estimation based scheduler (BLEST)~\cite{blest}, and ECF~\cite{ecf}. The details of our experiments and consequent results are discussed in Section \ref{sec:realexperiments}. All our results throughout evaluation are averaged over multiple runs.

\section{Simulation Setup and Results} 
\label{sec:simulation}

\begin{figure}[!t]
\centering
\begin{subfigure}{0.47\textwidth}
\centering
  \includegraphics[width=\textwidth]{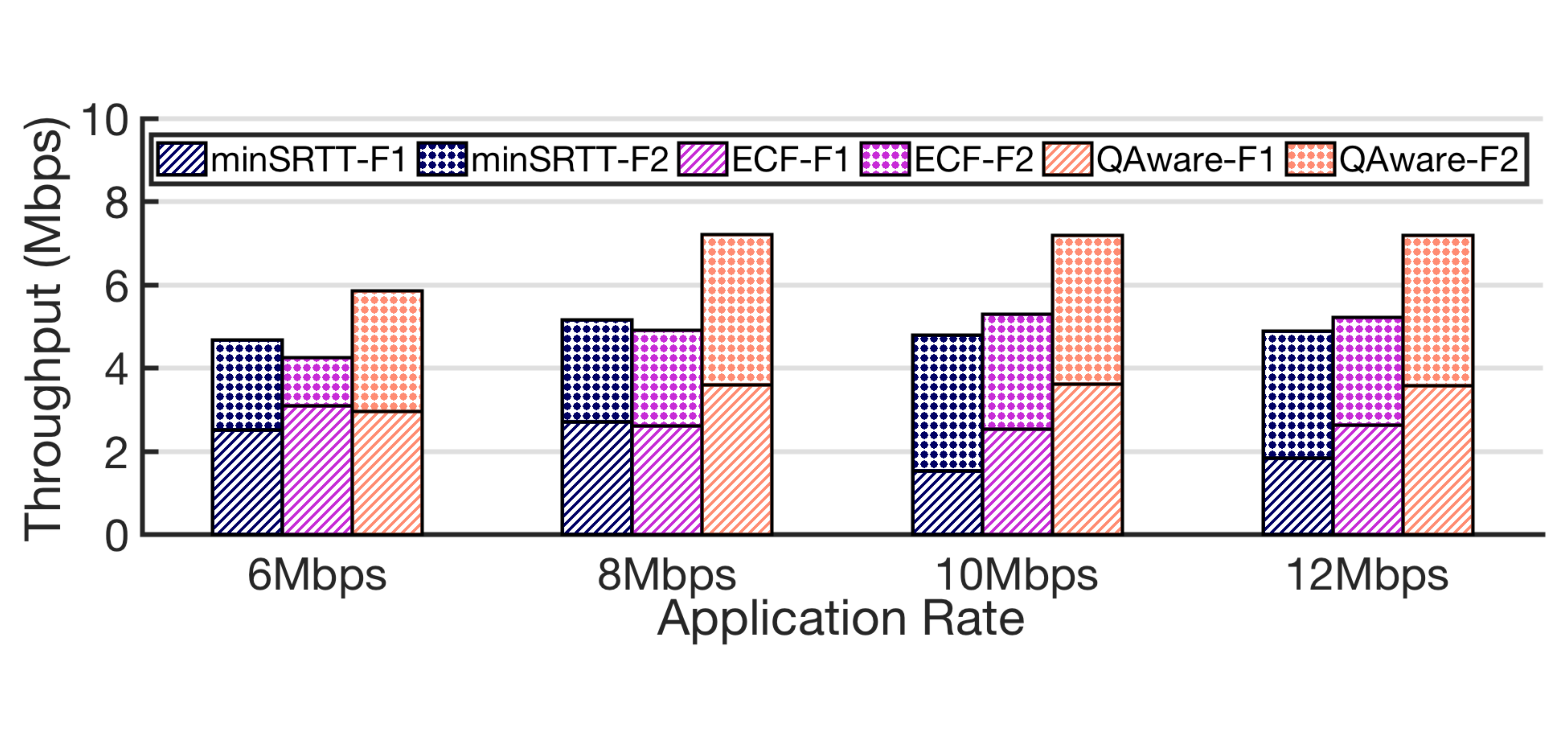}
  \caption{\label{fig:experiment1a_homo} Subflows F1 and F2 use links with PHY rates of $6$ Mbps.}
\end{subfigure}%
\vspace{0.005\textwidth}
\begin{subfigure}{0.47\textwidth}
\centering
  \includegraphics[width=\textwidth]{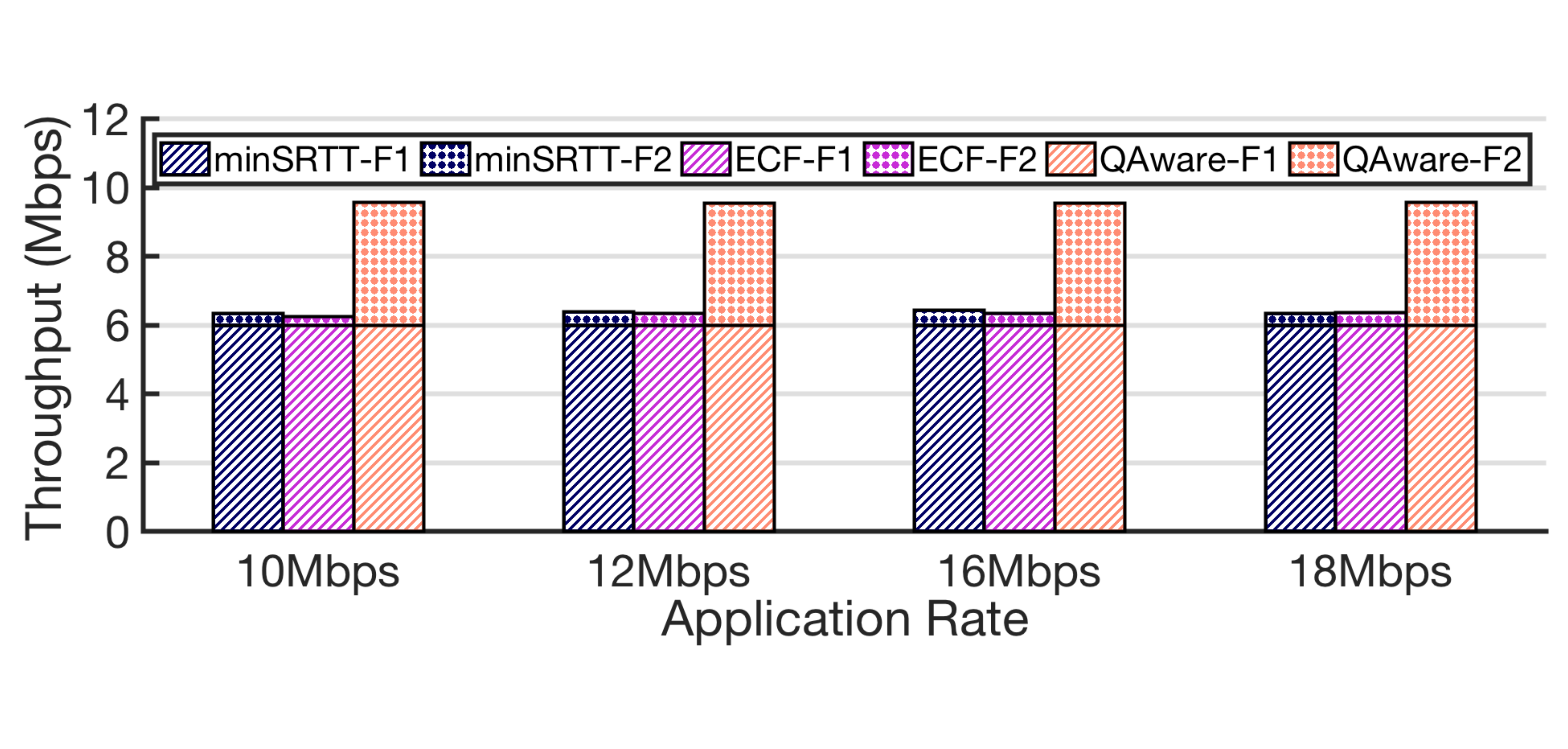}
  \caption{\label{fig:experiment1a_hetro} Subflow F1 and F2 use links with PHY rate of $12$ Mbps and $6$ Mbps respectively.}
\end{subfigure}
\caption{Throughput achieved by minSRTT, ECF and QAware schedulers for different CBR rates.}
\label{fig:throughput_no_error}
\end{figure}

We simulated network topologies of the kind shown in Figure~\ref{fig:netw_topology}. For all simulations, the links between the access points and the backbone switch and between the backbone switch and the server were modeled as wired links with rate $30$ Mbps and $50$ Mbps respectively. The client is connected to the two access points over wireless links with physical layer (PHY) rates in the range $6-12$ Mbps. These two wireless links provided the two network paths over which application data was sent. Both subflows use independent congestion control.

We simulated the following \emph{applications}: 
\begin{inparaenum}[i)]
\item constant bit rate (CBR) data from low to high rates, 
\item file transfer for sizes of $10-30$ MB, 
\item web browsing of top 10 out of the US Alexa-100 websites, and 
\item CBR with one of the paths being shared by UDP traffic. 
\end{inparaenum}
 For the applications, we simulated the following \emph{network configurations}: 
\begin{inparaenum}[i)]
\item both wireless links have the same rate, 
\item one link is much faster than the other, and 
\item one link drops TCP packets. 
\end{inparaenum}
Comparisons of QAware with ECF and minSRTT\footnote{In simulation, the scheduler assigns packets over independent TCP streams. We do not incorporate other MPTCP functionality such as re-transmission handler and path manager.}
demonstrate the benefits that are accrued by QAware because it optimally utilizes both network paths.



\begin{figure}[!t]
\centering
\begin{subfigure}{0.45\textwidth}
\centering
  \includegraphics[width=\textwidth]{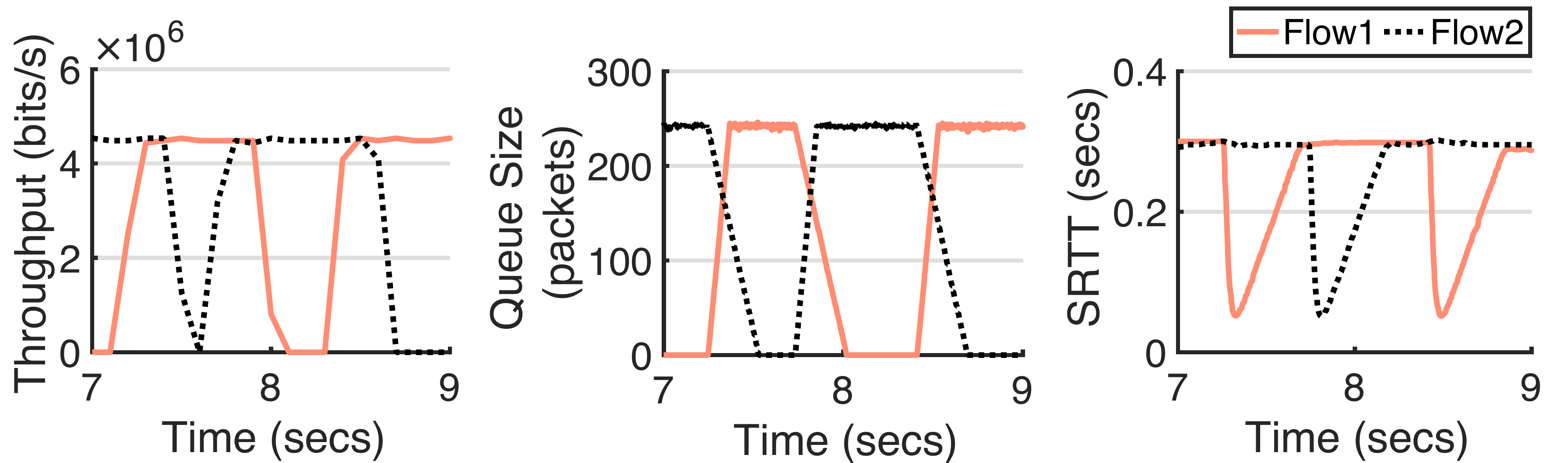}
  \caption{\label{fig:behavior_srtt}minSRTT Scheduler}
\end{subfigure}%
\vspace{0.005\textwidth}
\begin{subfigure}{0.45\textwidth}
\centering
  \includegraphics[width=\textwidth]{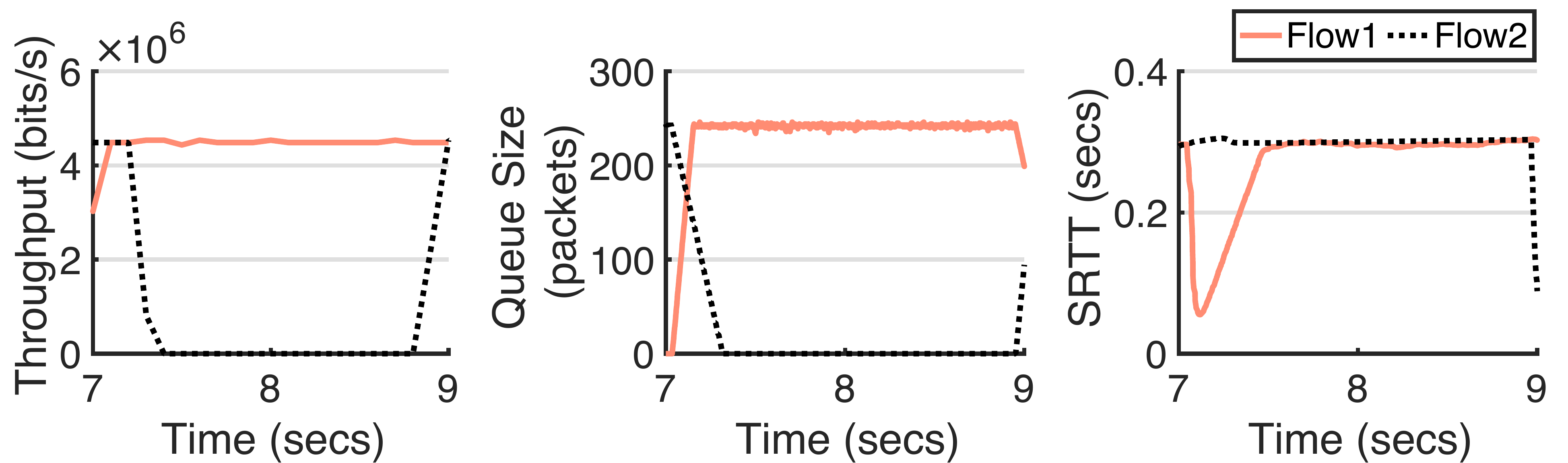}
  \caption{\label{fig:behavior_ecf}ECF Scheduler}
\end{subfigure}
\begin{subfigure}{0.45\textwidth}
\centering
  \includegraphics[width=\textwidth]{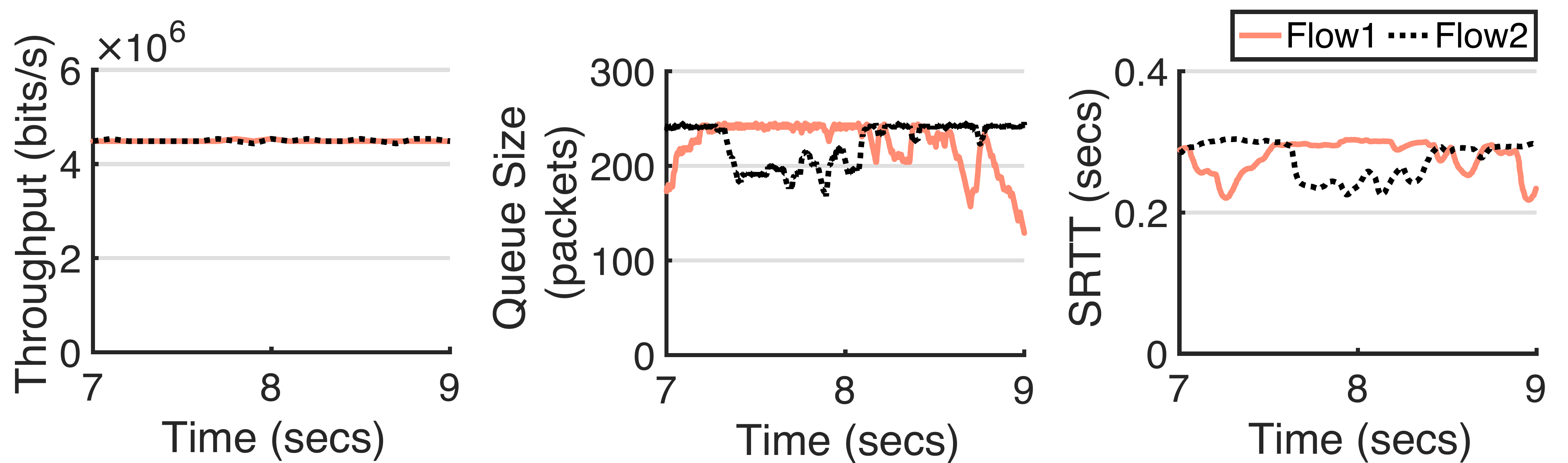}
  \caption{\label{fig:behavior_qa}QAware Scheduler}
\end{subfigure}%
\caption{Per-flow throughput, device driver queue occupancy, and SRTT behavior as a function of time. These correspond to the throughputs in Figure~\ref{fig:experiment1a_homo} and a CBR rate of $12$ Mbps.}
\label{fig:explanation}
\end{figure}


\subsection{Constant Bit Rate Traffic} \label{sec:cbrsimulation}
\subsubsection*{Access paths with no packet errors} 
Figure~\ref{fig:experiment1a_homo} shows the TCP throughputs obtained by the schedulers for increasing CBR rates. Each wireless link was configured with a PHY rate of $6$ Mbps. This results in homogeneous network paths. On average, QAware achieves percentage throughput gains of about $40\%$ over the rest. Further, note that all schedulers use both subflows. However, unlike the others, QAware utilizes both the subflows almost equally for the entire simulation time for all the CBR loads. To better understand their behaviors, consider Figure~\ref{fig:explanation}, which shows for each scheduler and subflow, the variation of throughput, device driver queue occupancy, and smoothed RTT, as a function of time, for a $2$ second interval. The CBR rate was set at $12$ Mbps. From the subflow throughputs and queue occupancy, it is clear that QAware uses both subflows almost simultaneously. ECF uses just one subflow for most of the interval, and while minSRTT uses both flows during the interval, it switches between them very infrequently. Both minSRTT and ECF rely on the delayed feedback provided by SRTT and so end up scheduling packets to one subflow for longer intervals than QAware. Essentially, they switch flows when SRTT of the subflow in use exceeds that of the other subflow. In addition, ECF, by design, declines scheduling opportunities to a subflow with a larger RTT and prefers to wait for faster subflows. This explains the reason for using one flow for a longer duration than minSRTT scheduler. In minSRTT and ECF, subflows experience swings in SRTT. The SRTT increases linearly while it is the subflow of choice. This increase eventually makes the subflow less desirable than the other and the scheduler switches to the other flow, which, due to the current low occupancy in the corresponding device queue, experiences low SRTT.\footnote{Our observations with respect to QAware and minSRTT for three homogeneous paths are similar. We skip them due to lack of space.}



\begin{figure}[!t]
\centering
  \includegraphics[width=0.47\textwidth]{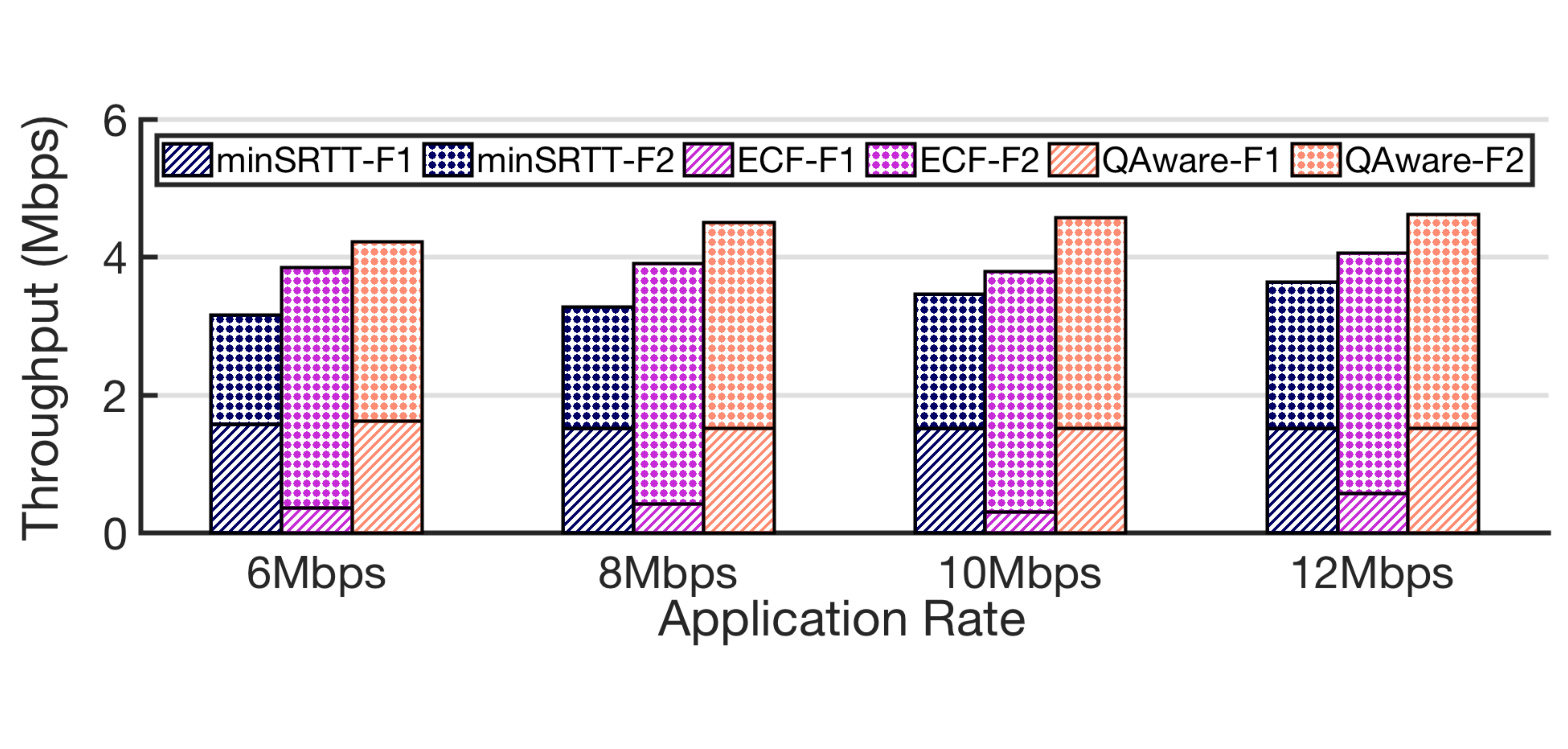}
\caption{Per-flow throughput comparison for different CBR rates where subflow F1 experiences a packet drop rate of $10^{-2}$.}
\label{fig:experiment1b_homo}
\end{figure}

Figure~\ref{fig:experiment1a_hetro} shows throughputs obtained by the CBR application when the PHY rate of one of the wireless links is $6$ Mbps and the other is $12$ Mbps. While all schedulers utilize the subflow using the $12$ Mbps link equally, QAware also utilizes the subflow mapped on the $6$ Mbps link. On average, QAware achieves throughput gains of about $50\%$ over the rest.

\subsubsection*{Access paths with packet errors} 
Figure~\ref{fig:experiment1b_homo} shows the throughput obtained when one subflow suffers a packet loss rate of about $10^{-2}$. Both wireless links have PHY rates of $6$ Mbps. Upon detecting packet loss, the congestion window of the subflow decreases based on \textit{TCP congestion avoidance algorithm}, which limits the number of packets that can be sent on that subflow. Even in this situation, QAware is able to exploit both subflows better and achieves about $32\%$ and $15\%$ improvement over minSRTT and ECF respectively. For the case when the wireless links are $12$ Mbps and $6$ Mbps with an error on the slower link,  the corresponding gains are $53\%$ and $6\%$ (figure not shown due to space limitations). Note that since ECF is biased toward using the faster path, it performs almost as well as QAware when the error-free path has a faster wireless link. On the other hand, while minSRTT uses the error-prone path better than ECF, it is unable to make good use of the error-free path as the other two schedulers.

\subsection{Fixed Size File Transfer} 
\begin{figure}[!t]
\centering
  \includegraphics[width=0.4\textwidth]{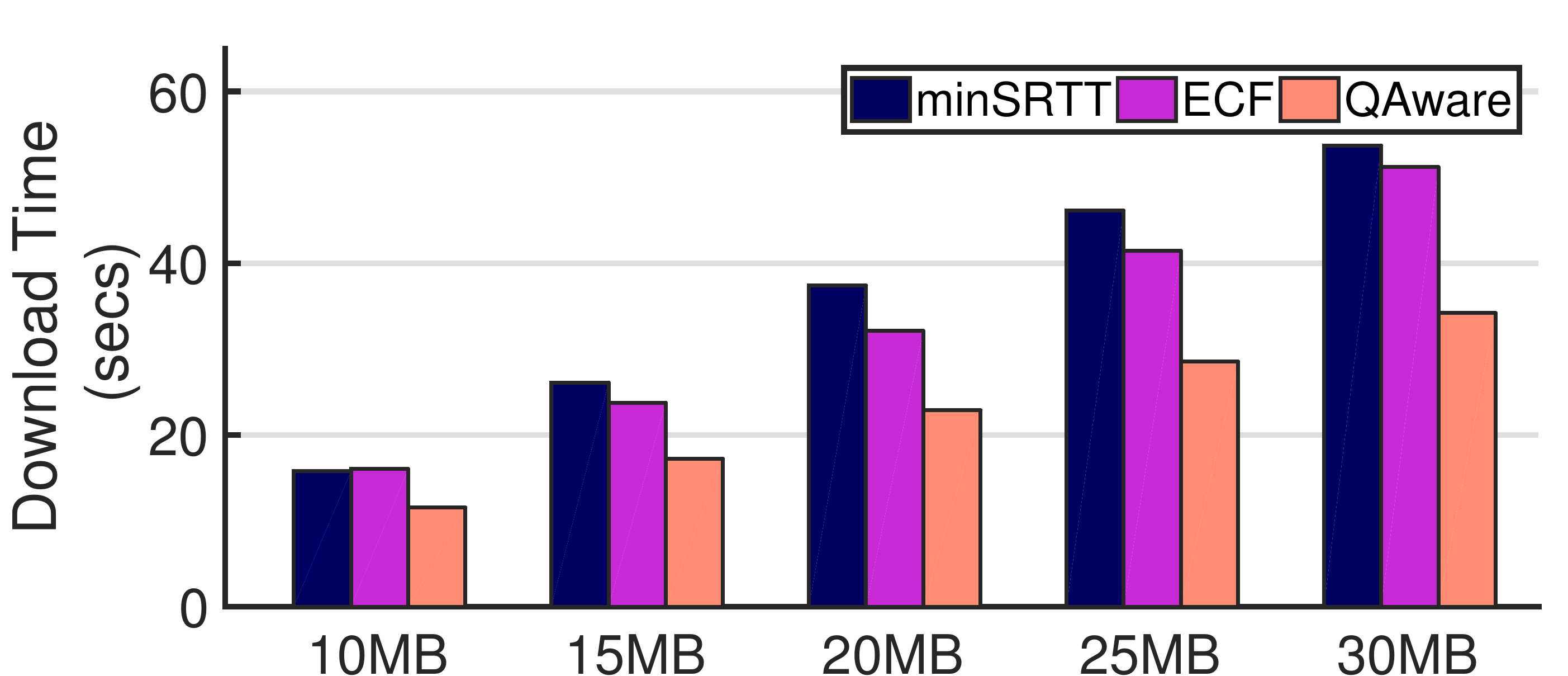}
\caption{File download completion times when both subflows use wireless link with PHY rate of $6$ Mbps.}
\label{fig:download_simulator}
\end{figure}
Figure~\ref{fig:download_simulator} shows the download completion time achieved by the three schedulers for five different file sizes ranging from 10MB to 30MB. Both wireless links were set to a PHY rate of $6$ Mbps. Observe that QAware obtains the least download time for all the file sizes. This is explained by its ability to effectively utilize both the subflows for data transfer. The performance gap increases proportionally with file size. Overall, QAware achieves $35\%$ and $30\%$ reduction in average download time over minSRTT and ECF respectively.

\subsection{Web-browsing}

\begin{table}[!b]
\centering
\begin{tabular}{c|ccccc} 
\toprule
\textbf{Website}   & News   & Tech   & Radio  & Shopping & Finance  \\ 
\midrule
\textbf{\#Objects} & 202    & 67     & 66.2   & 52.2     & 39.7     \\
\textbf{Size (KB)} & 3821.2 & 2152.2 & 2453   & 1000.7   & 1988.1   \\ 
\hline\midrule
\textbf{Website}   & Wiki   & Market & Social & Movie    & Travel   \\ 
\midrule
\textbf{\#Object}  & 28     & 49     & 69     & 39       & 21       \\
\textbf{Size (KB)}      & 601.2  & 2032.8 & 1700.2 & 845.7    & 2000.4   \\
\bottomrule
\end{tabular}
\caption{Web objects for traffic generation}
\label{table:webbrowsing}
\end{table}


\begin{figure}[!t]
\centering
  \includegraphics[width=0.48\textwidth]{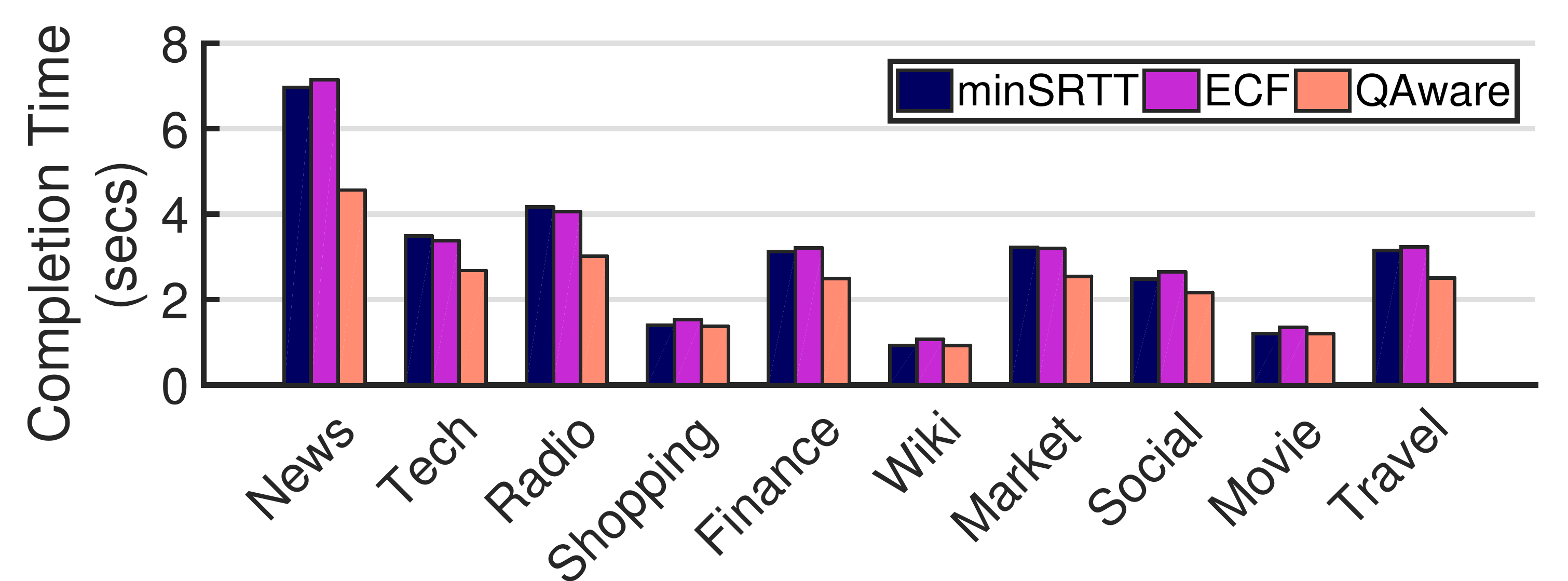}
\caption{Download completion time for 10 websites from top U.S. Alexa-100 websites.}
\label{fig:web_homo}
\end{figure}

To simulate web browsing, we deployed objects of 10 out of top U.S. Alexa-100 websites, which are summarized in Table \ref{table:webbrowsing}, in our simulated server. The client consecutively downloaded relevant objects of each website from the server at a variable rate between 10Mbps to 30Mbps chosen in a probabilistic manner. We compared scheduler performance for when both wireless links are $6$ Mbps and when one of the links is $12$ Mbps. QAware achieves a significant reduction in download completion time for both configurations, specifically up to $35\%$ for the former (see Figure~\ref{fig:web_homo}) and up to $28\%$ for the latter (figure not shown due to space limitations). On the other hand, ECF and minSRTT perform similarly. 


\subsection{Multiple Applications}

\begin{figure}[!t]
\centering
  \includegraphics[width=0.42\textwidth]{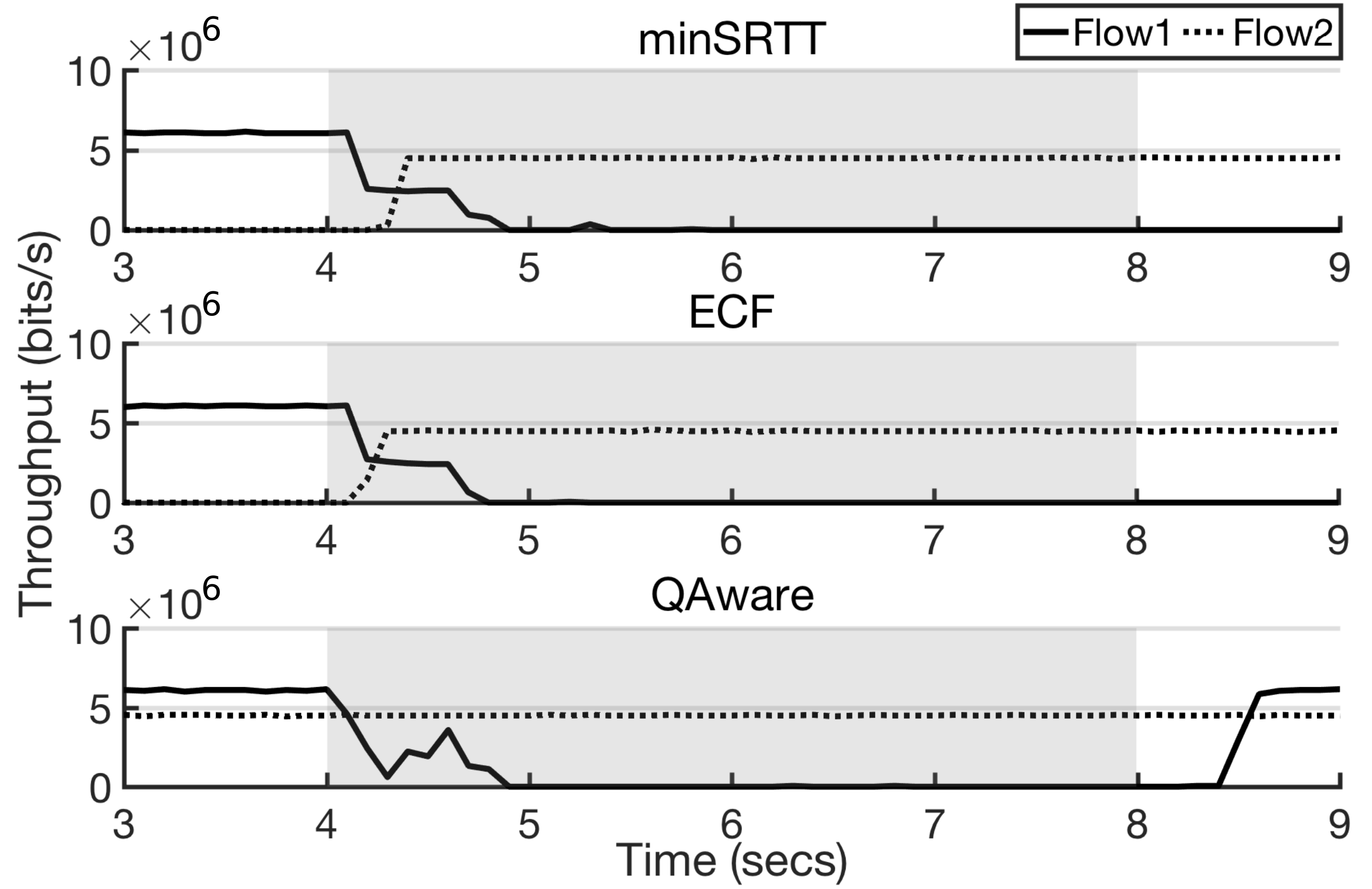}
\caption{Per-flow throughputs when the interface used by subflow F1 sees UDP traffic for $4$ seconds (greyed).}
\label{fig:mixed_traffic}
\end{figure}

In current computing environments, end hosts typically run multiple applications which must share the interfaces available at the host for network transfers. An ideal MPTCP scheduler must be able to efficiently adapt to bandwidth competition on bottleneck links in such coexisting environment. To evaluate the impact of such sharing on the schedulers, we used the following setup. The PHY rates of the wireless links were set to $9$ and $6$ Mbps. A CBR application generated data for a 10 second interval and used both the MPTCP subflows. The results are shown in Figure \ref{fig:mixed_traffic}. 

Starting at 4 seconds, we introduced traffic from a UDP application that used the network path with the $9$ Mbps wireless link. The greyed area in the figure denotes the time duration when both MPTCP and UDP applications were active at the client. The UDP traffic lasted for 4 seconds. Before the start of the UDP traffic, only QAware scheduler was utilizing both available subflows. Once the UDP application starts, the device queue of the $9$ Mbps wireless link saturates. QAware, however, quickly adapts to it and reduces the traffic being sent on the corresponding subflow. All the while, it keeps utilizing the subflow over the slower wireless link. On the other hand, both minSRTT and ECF need to wait for several RTT updates for the impact of UDP traffic on queue wait times to get reflected in the SRTT of the subflow. Lastly, unlike the other schedulers, QAware is also quick to detect the availability of the subflow after the 8 second mark, which is when the UDP application stops its transfer. Overall, QAware leads to gains of about $40\%$ over minSRTT and about $50\%$ over ECF.

\section{Real-World Setup And Experiments}
\label{sec:realexperiments}
\begin{figure}[!t]             
\begin{center}
\includegraphics[width=0.38\textwidth]{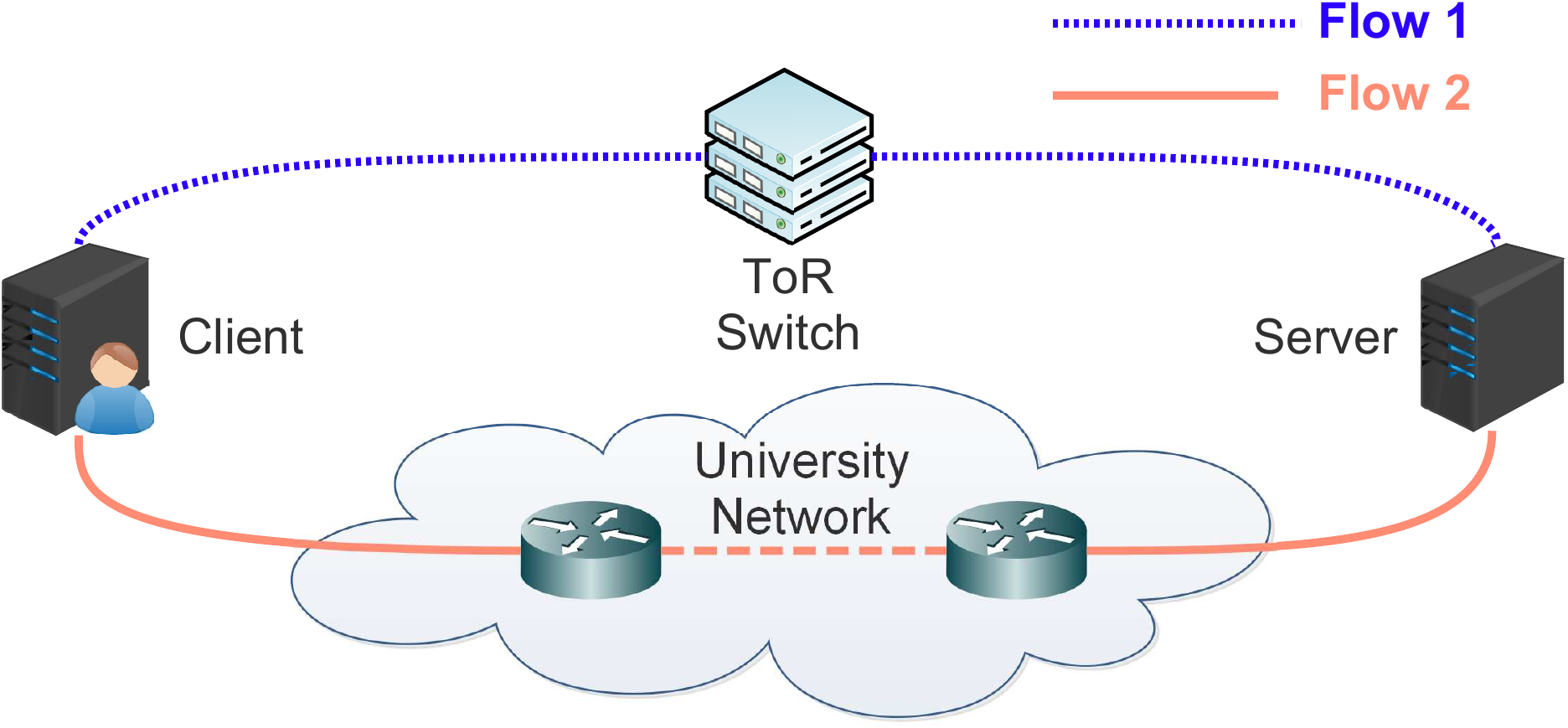}
\caption{Real network testbed in university datacenter.}
\label{fig:realsetup}
\end{center}
\end{figure}

We next examine QAware's performance in real network environments. Figure \ref{fig:realsetup} shows our test network topology in University of Helsinki data center. We assign two similar machines with 16 core AMD Opteron processor, 8 GB DDR2 RAM running Ubuntu 16.04 LTS with latest stable MPTCP implementation (version 0.93, based on Linux kernel v4.9.60 \cite{mptcplinuximpl}) as client and server. The implementation uses default congestion control algorithm (coupled OLIA). Both machines are interconnected via two separate Gigabit Ethernet interfaces. One Ethernet connection is routed through internal University of Helsinki network and therefore encounters background traffic from University staff. It has an end-to-end RTT of $>$1ms. The other connection is over Top-of-Rack (ToR) switch with RTT $<$1ms.

We compare QAware with the following schedulers: 
\begin{inparaenum}[i)]
\item minSRTT, 
\item Delay Aware Packet Scheduler (DAPS)~\cite{daps}
\item Blocking Estimation based Scheduler (BLEST)~\cite{blest}, and 
\item Earliest Completion First (ECF)~\cite{ecf}\footnote{For DAPS and BLEST, we use the implementation at \url{https://bitbucket.org/blest_mptcp/nicta_mptcp}. For ECF, we use the implementation at \url{http://cs.umass.edu/~ylim/mptcp_ecf}} \footnote{DAPS, BLEST, and ECF are implemented on MPTCP v0.89 whereas the default minSRTT and QAware are based on MPTCP v0.93. We could not implement QAware on MPTCP v0.89 as it is based on Linux v3.18 which does not support BQL. Please see \cite{mptcplinuximpl} for exact changes between the two versions.}.
\end{inparaenum}
We first compare scheduler performance for application generating bulk traffic. This workload provides a qualitative validation of the results we obtained in Section~\ref{sec:simulation}. We further present scheduler performance for DASH video streaming and web file downloads. We used the Linux Traffic Control system (\emph{tc}) in combination with a Hierarchical Token Bucket (HTB) packet scheduler using Statistical Fair Queuing (SFQ) for network shaping. In between runs, we flushed out the TCP cache to ensure that each run is independent of the next. All our results are averaged over ten runs. 

\subsection{Bulk Traffic}

In this section, we compare QAware's performance with other schedulers for high application transfer rate over both subflows. We performed experiments with different settings of delays along the two paths. The setting includes 
\begin{inparaenum}[i)]
\item default path delays ($< 1$ms and $> 1$ms),
\item delay shaping to introduce $40$ms of delay along one path and $80$ms along the other, and 
\item $40$ms along one path and $160$ms along the other. 
\end{inparaenum}
Path bandwidths corresponding to the different delays are stated in Table~\ref{table:netwhetro}.

\begin{figure}[!t]
\centering
\begin{subfigure}{0.165\textwidth}
\centering
  \includegraphics[width=\textwidth]{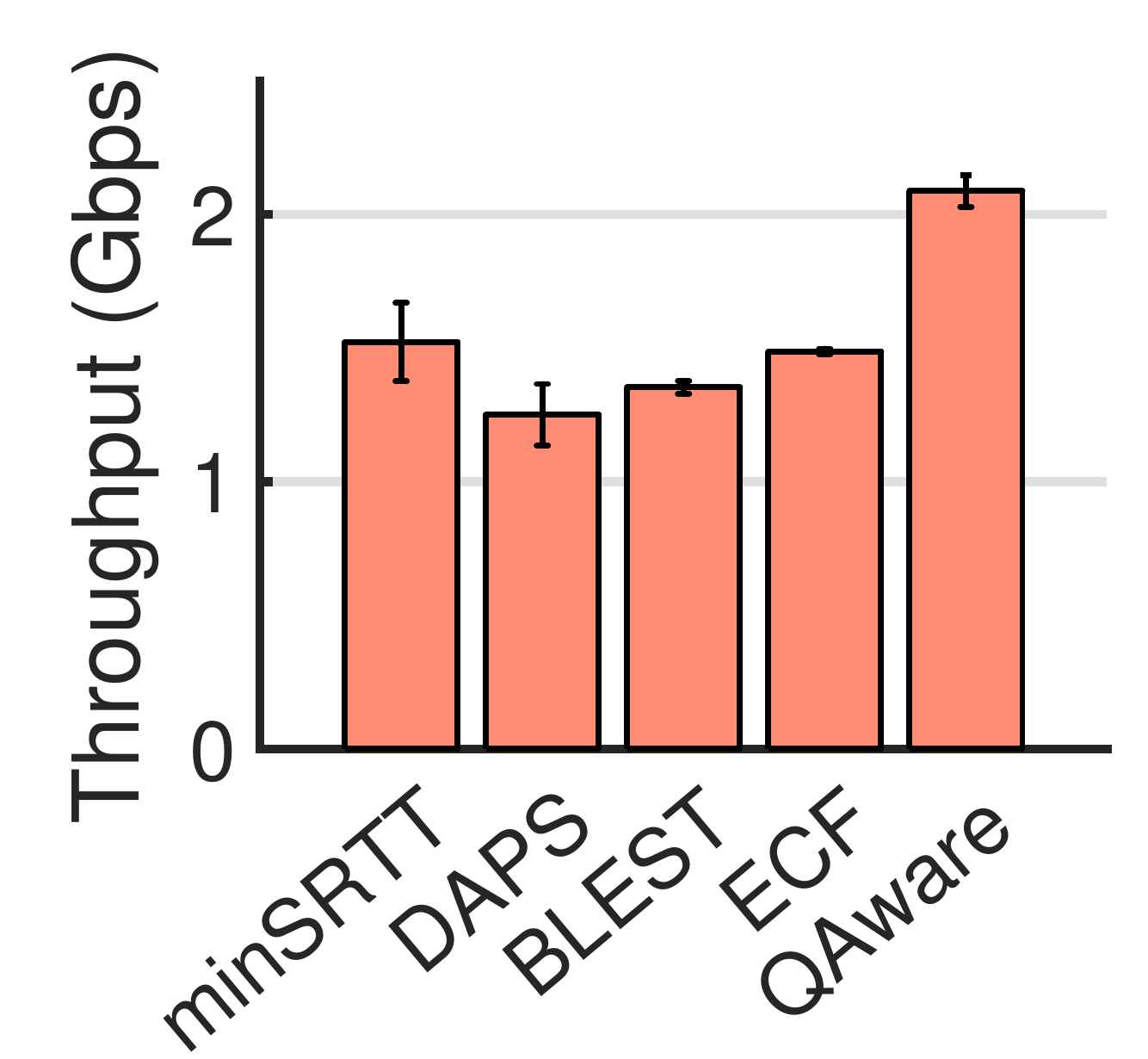}
  \caption{\label{fig:iperf_linux}Default}
\end{subfigure}%
\hfill
\begin{subfigure}{0.16\textwidth}
\centering
  \includegraphics[width=\textwidth]{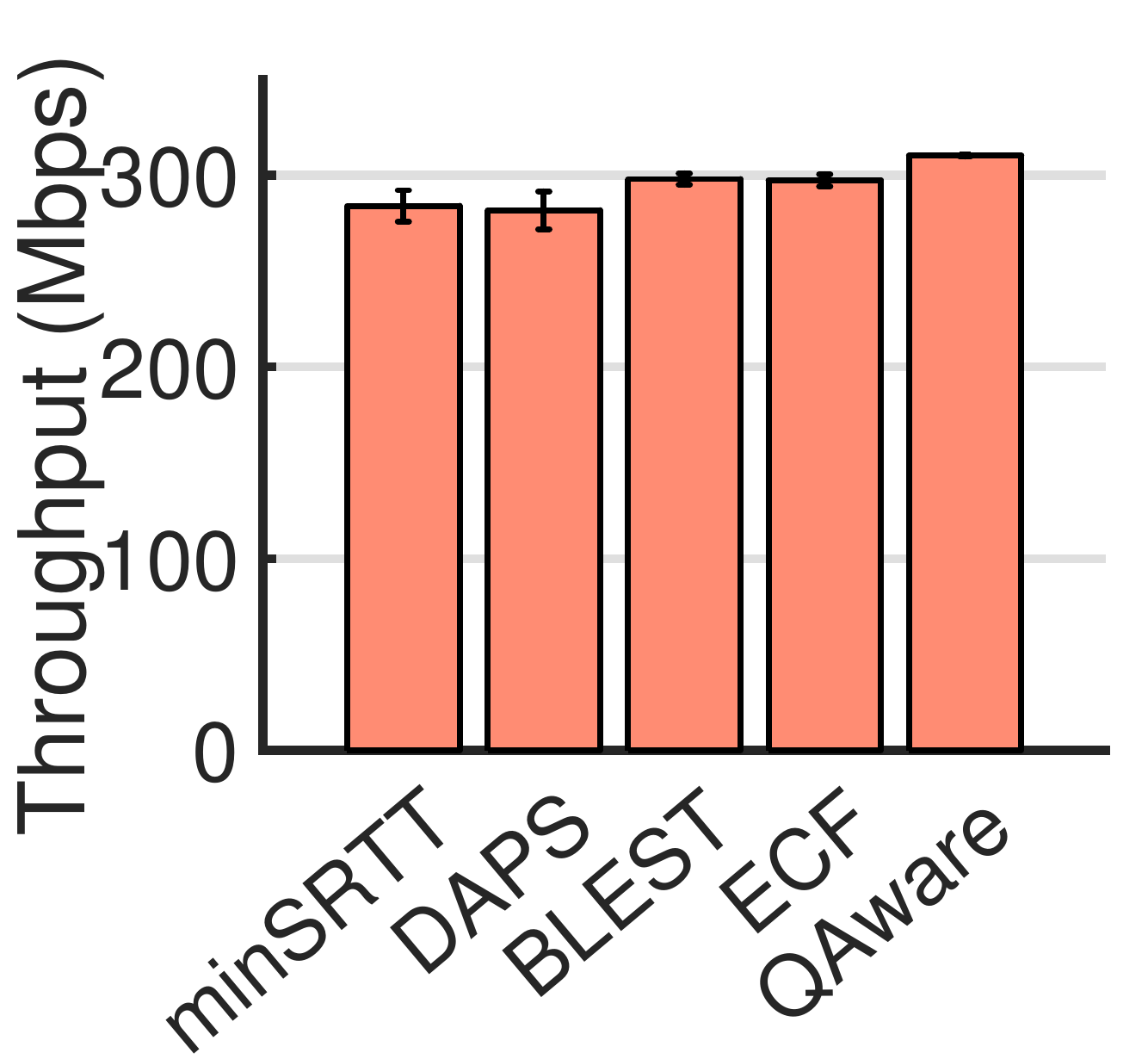}
  \caption{\label{fig:delay-1_2}40+80ms}
\end{subfigure}%
\hfill
\begin{subfigure}{0.16\textwidth}
\centering
  \includegraphics[width=\textwidth]{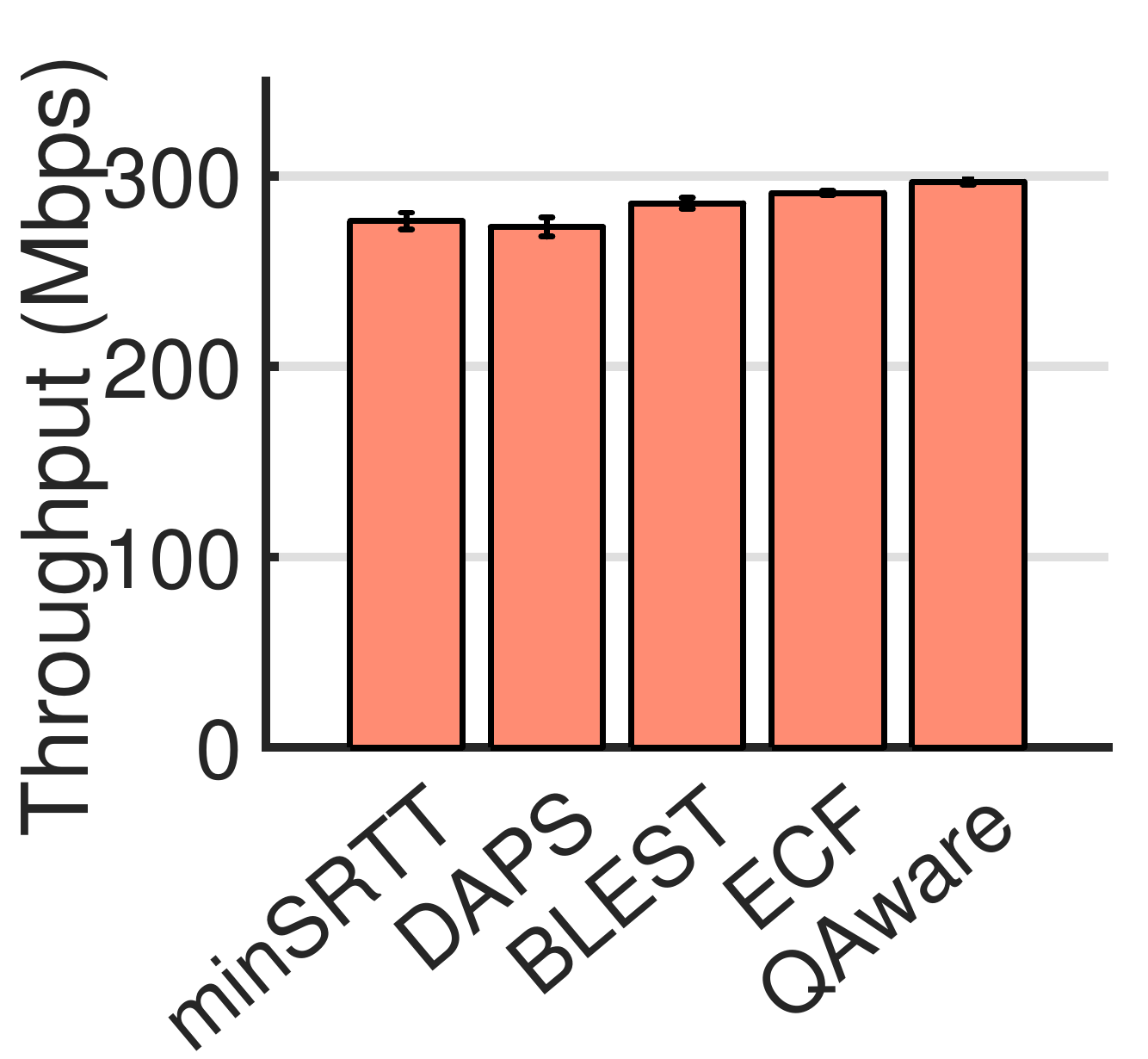}
  \caption{\label{fig:delay-1_4}40+160ms}
\end{subfigure}
\caption{Bulk Traffic throughputs for different access path delays.}
\label{fig:delay_linux}
\end{figure}

Figure~\ref{fig:iperf_linux} compares average throughput obtained by different schedulers for default path delays. QAware achieves more than 45\% increase in throughput compared to DAPS, BLEST and ECF. QAware also provides an improvement of $37\%$ over the default minSRTT scheduler. Interestingly, the minSRTT scheduler outperforms DAPS, BLEST, and ECF in the experiment. We attribute minSRTT's efficiency to two reasons. Firstly, DAPS, BLEST and ECF schedulers have been designed to improve MPTCP performance for heterogeneous delays along available network paths. In fact, BLEST and ECF even go as far as not sending an available packet on a slower subflow and wait for the faster subflow to become available. When subflows witness similar delays (as in the current case), the default scheduler places more packets on each path as opposed to DAPS, BLEST, and ECF. Secondly, based on latest MPTCP kernel, minSRTT enjoys several code improvements and optimizations.

For when the path delays are $40$ and $80$ms, QAware yields an average throughput of 310 Mbps which is an improvement of about $10\%$ over the default scheduler and DAPS and $5\%$ over ECF and BLEST (shown in Figure \ref{fig:delay-1_2}). As presented in Figure \ref{fig:delay-1_4}, all schedulers perform quite similar to each other as all try to fully utilize the lower delay subflow when path delays are $40$ and $160$ms.  In this case, QAware still manages to achieve an improvement of about $7\%$ over the default scheduler and DAPS, and about $4\%$ over BLEST and ECF. 

\subsection{Video Streaming}

\begin{figure}[!t]    
\centering  
\begin{subfigure}{0.158\textwidth}
\centering
  \includegraphics[width=\textwidth]{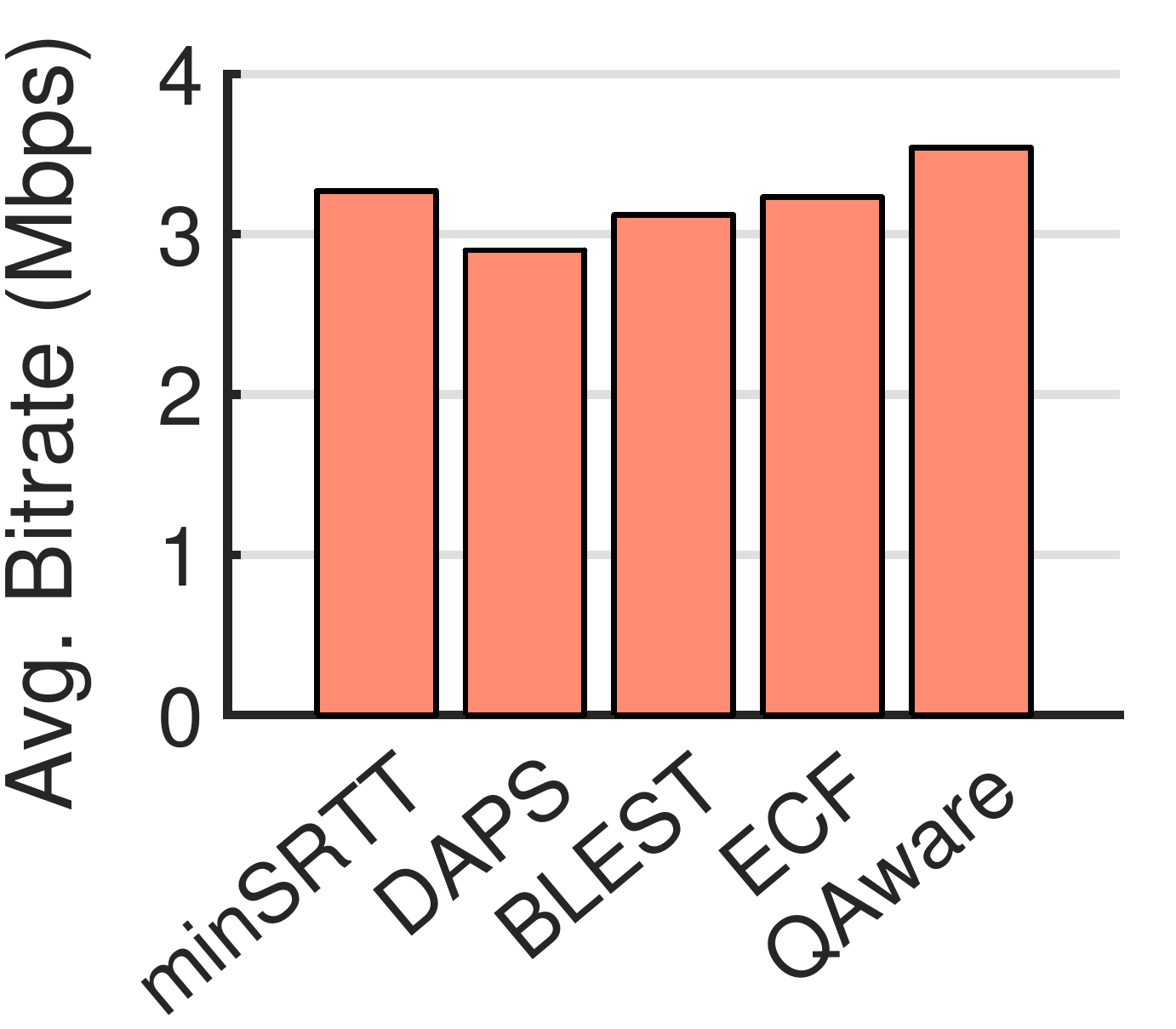}
  \caption{\label{fig:video-2+2}2+2 Mbps}
\end{subfigure}
\hfill
\begin{subfigure}{0.158\textwidth}
\centering
  \includegraphics[width=\textwidth]{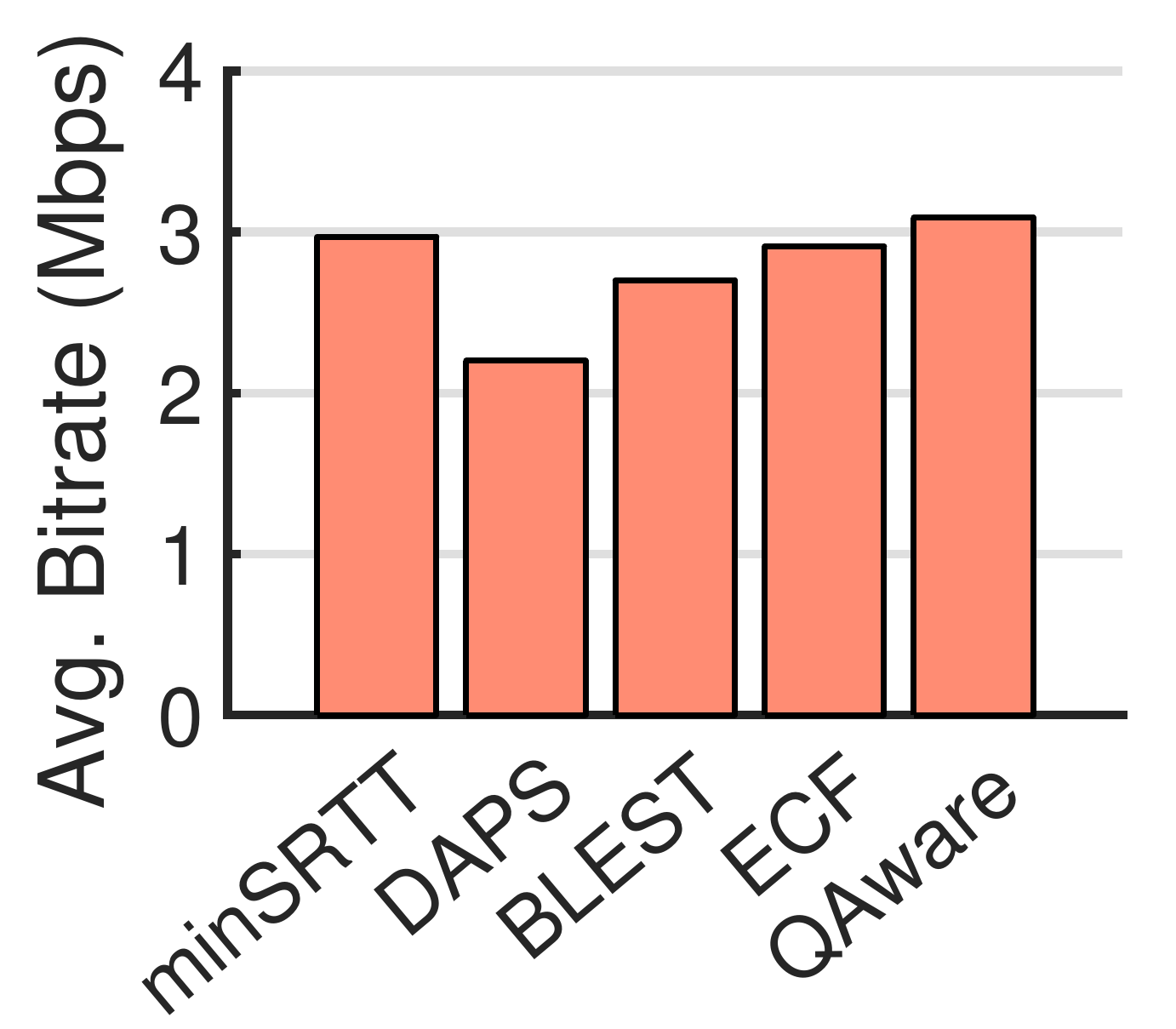}
  \caption{\label{fig:video-2+16}2+1.6 Mbps}
\end{subfigure}
\hfill
\begin{subfigure}{0.158\textwidth}
\centering
  \includegraphics[width=\textwidth]{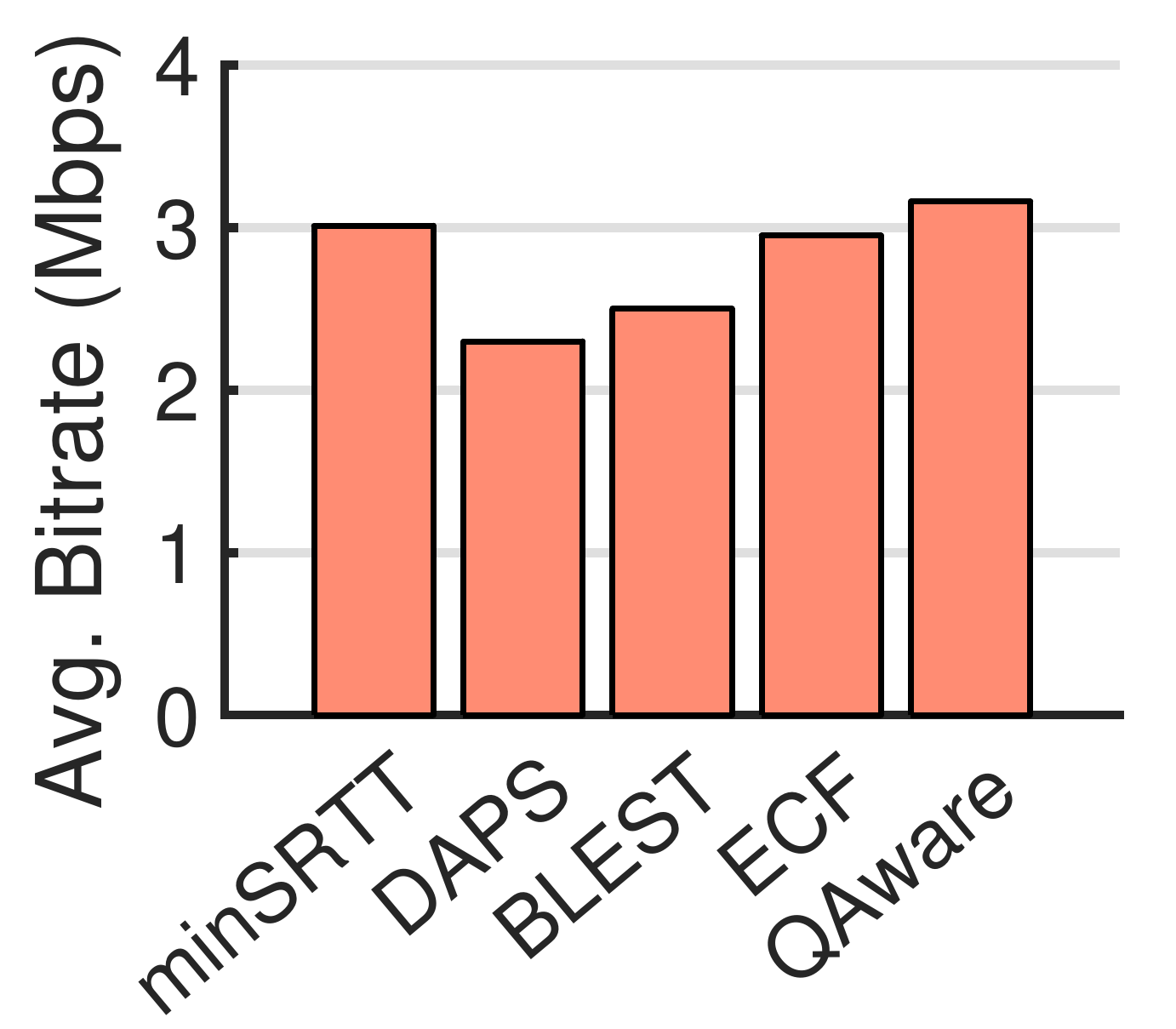}
  \caption{\label{fig:video-24+16}2.4 +1.6 Mbps}
\end{subfigure}
\caption{Average bitrate in video streaming for different path bandwidths.}
\label{fig:videoexp}  
\end{figure}

\begin{table}[!ht]
\centering

\begin{subtable}{\linewidth}
\centering
\begin{tabular}{lllll} 
\toprule
Delay (ms)       & 1   & 40& 80& 160    \\
Bandwidth (Mbps) & 950 & 600& 300& 200  \\
\bottomrule
\end{tabular}
\caption{Configurations for Bulk Traffic Experiments}
\label{table:netwhetro}
\end{subtable}

\vspace*{1em}

\begin{subtable}{\linewidth}
\centering
\begin{tabular}{llll} 
\toprule
Bandwidth (Mbps) & 2.4 & 2& 1.6  \\
Delay (ms)       & 10  & 20& 30  \\
\bottomrule
\end{tabular}
\caption{Configurations for Video Streaming Experiments}
\label{table:netwvideo}
\end{subtable}

\caption{\ref{table:netwhetro} shows bandwidth achieved by delay throttling on a 1Gbps Ethernet interface whereas \ref{table:netwvideo} presents values after both bandwidth and delay shaping}
\end{table}

Streaming is a dominant Internet use case and is widely adopted by content providers such as Netflix and YouTube~\cite{streamingapp}. We set up a DASH server and host \textit{Big Buck Bunny}, available from a public dataset, on it \cite{dashDataset}. We configured the streaming server to provide five representations of the video from 240p to 1080p (same as most content providers).  We re-encoded each representation in at least three different bitrates with overall available bit rates from 128Kbps to 3.8Mbps using H.264/MPEG-4 AVC codec. The streaming client employs an Adaptive Bit Rate (ABR) algorithm to download video segments according to the available network bandwidth. We throttled our testbed bandwidth to match the bitrates of DASH encodings. Table \ref{table:netwvideo} shows the average delay measured at client-side for each bandwidth configuration. We evaluate and compare QAware's performance with other schedulers for when the two subflows 
\begin{inparaenum}[i)]
\item have bandwidths of 2 Mbps, 
\item have bandwidths of 2 Mbps and 1.6 Mbps, and 
\item have bandwidths of 2.4 Mbps and 1.6 Mbps. 
\end{inparaenum}

From Figure \ref{fig:videoexp}, we observe that QAware improves the performance of streaming applications in all network conditions. The performance improvement is more significant in scenarios where the path bandwidths are similar (8\% and 5\% with respect to default and 10\% and 6\% with respect to ECF, in Figures~\ref{fig:video-2+2} and \ref{fig:video-2+16} respectively) as QAware utilizes available paths more efficiently than other schedulers. DAPS consistently gives the worst performance out of all schedulers due to its strong dependence on RTT ratio of two subflows. 

\subsection{Web File Download}
\begin{figure}[!b]             
\begin{center}
\includegraphics[width=0.42\textwidth]{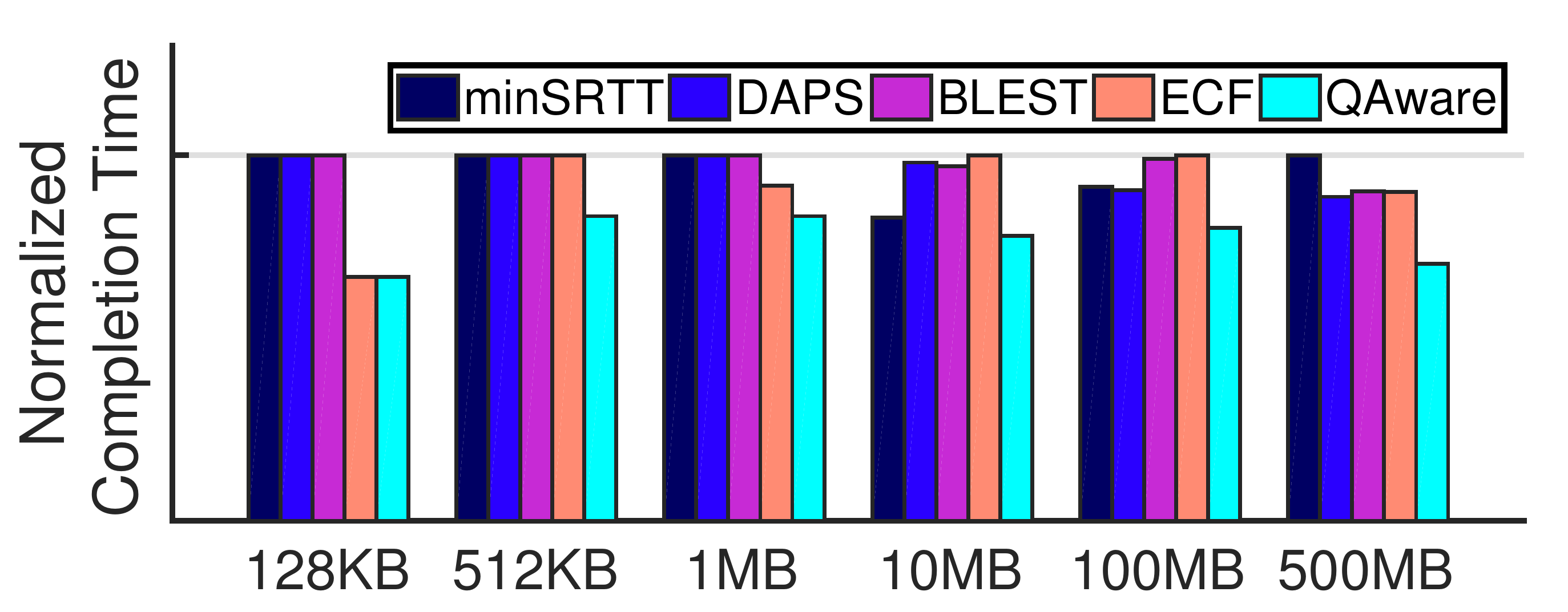}
\caption{Normalized download completion time for different file sizes ({smaller is better}).}
\label{fig:linux-file-download}
\end{center}
\end{figure}

We now evaluate QAware's performance for simple web downloads using \emph{curl}. We set up an HTTP server using Apache 2.2.22 and hosted varying file sizes of range 128KB to 500MB. We eliminate application connection time by only considering the transport-level time in overall download completion time observed at the client. Figure \ref{fig:linux-file-download} presents the average completion time normalized to the maximum achieved value by scheduler for a given file size.

For small web transfers ($<$1MB) all schedulers perform quite similar to each other (it took 0.002s to download a 128 KB file by QAware vs. 0.003s by minSRTT). This is because for small data transfers, the bandwidth of the primary subflow is more than capable of single shot transmission and thus MPTCP rarely switches to the secondary subflow. Therefore, until the performance of primary subflow degrades during transfer, the choice of the scheduler does not affect the performance for small files. The default and DAPS scheduler achieve lower completion time for medium file sizes ($\approx$10/100 MB) in comparison to BLEST and ECF. This is likely because BLEST and ECF add additional delays by waiting for the faster subflow to become available. For large files (500 MB), BLEST and ECF utilize faster subflow more efficiently than default and DAPS, thus achieving a lower completion time. QAware always outperforms other schedulers and realizes up to 20\% decrease in completion time for medium file sizes (0.709s by QAware vs. 0.895s by ECF for 100 MB file) and 30\% for large file downloads (3.46s by QAware vs. 4.93s by minSRTT for 500 MB).

\section{Conclusion}\label{sec:conclusion}




We proposed, QAware, a novel cross-layer MPTCP scheduler that combines hardware device queue occupancy and TCP RTT for efficient scheduling decisions. We detailed its design and implementation. We evaluated QAware using an extensive set of simulations and real network experiments for various network configurations and applications such as bulk data transfers, web browsing, web file downloads, and video streaming. Comparisons with various state-of-the-art schedulers such as DAPS, BLEST, and ECF were used to demonstrate the efficacy of QAware. It outperformed other schedulers in all network configurations and workloads we tested. 
Further, we show that QAware quickly adapts to co-existing applications and sudden variations in network conditions. We have open-sourced QAware's implementation as a modular scheduler for latest stable MPTCP Linux release.

\section*{Acknowledgment}

This research was funded by TCS Research Scholarship Program, EU FP7 Marie Curie Actions Cleansky Project (Contract No. 607584) and Young Faculty Research Fellowship (Visvesvaraya Ph.D. scheme) from MeitY, Govt. of India.

\bibliographystyle{abbrv} 
\bibliography{IEEEtran}

\begin{thebibliography}{10}

\bibitem{streamingapp}
{Global Internet phenomenon}.
\newblock \url{www.sandvine.com/.../global-internet-
  phenomena-report-latin-america-and-north-america}.

\bibitem{applemptcp}
{Apple Inc.}
\newblock {Use Multipath TCP to create backup connections for iOS}.
\newblock \url{https://support.apple.com/en-us/HT201373}, 2017.

\bibitem{slow-path-adaptation}
S.~H. Baidya and R.~Prakash.
\newblock Improving the performance of multipath tcp over heterogeneous paths
  using slow path adaptation.
\newblock In {\em 2014 IEEE International Conference on Communications (ICC)},
  2014.

\bibitem{makewififast}
{Bufferbloat community}.
\newblock {Make WiFi fast project}.
\newblock \url{https://www.bufferbloat.net/projects/make-wifi-fast/wiki/},
  2014.

\bibitem{bqldrivers}
{Bufferbloat community}.
\newblock {The FlowQueue-CoDel Packet Scheduler and Active Queue Management
  Algorithm}.
\newblock \url{https://tools.ietf.org/id/draft-ietf-aqm-fq-codel-06.html},
  2014.

\bibitem{receiver-scheduler}
Y.~Cao, Q.~Liu, G.~Luo, and M.~Huang.
\newblock Receiver-driven multipath data scheduling strategy for in-order
  arriving in sctp-based heterogeneous wireless networks.
\newblock In {\em 2015 IEEE 26th Annual International Symposium on Personal,
  Indoor, and Mobile Radio Communications (PIMRC)}, 2015.

\bibitem{crosslayer_video}
X.~Corbillon, R.~Aparicio-Pardo, N.~Kuhn, G.~Texier, and G.~Simon.
\newblock Cross-layer scheduler for video streaming over mptcp.
\newblock In {\em Proceedings of the 7th International Conference on Multimedia
  Systems}, MMSys '16.

\bibitem{blest}
S.~Ferlin, Ã.~Alay, O.~Mehani, and R.~Boreli.
\newblock Blest: Blocking estimation-based mptcp scheduler for heterogeneous
  networks.
\newblock In {\em 2016 IFIP Networking Conference (IFIP Networking) and
  Workshops}, 2016.

\bibitem{dems}
Y.~E. Guo, A.~Nikravesh, Z.~M. Mao, F.~Qian, and S.~Sen.
\newblock Accelerating multipath transport through balanced subflow completion.
\newblock In {\em Proceedings of the 23rd Annual International Conference on
  Mobile Computing and Networking}, MobiCom '17, 2017.

\bibitem{mp-dash}
B.~Han, F.~Qian, L.~Ji, and V.~Gopalakrishnan.
\newblock Mp-dash: Adaptive video streaming over preference-aware multipath.
\newblock In {\em Proceedings of the 12th International on CoNEXT}, 2016.

\bibitem{packet-scheduling}
J.~Hwang and J.~Yoo.
\newblock Packet scheduling for multipath tcp.
\newblock In {\em Seventh International Conference on Ubiquitous and Future
  Networks}, 2015.

\bibitem{mptcplinuximpl}
M.~T. IETF.
\newblock {MPTCP Linux implmentation v0.93}.
\newblock \url{http://multipath-tcp.org/pmwiki.php?n=Main.Release93}, 2017.

\bibitem{mptcprfc}
{Internet Engineering Task Force (IETF)}.
\newblock {Architectural Guidelines for Multipath TCP Development}.
\newblock \url{https://tools.ietf.org/html/rfc6182}, 2011.

\bibitem{ietfbql}
{Internet Engineering Task Force (IETF)}.
\newblock {BQL enabled drivers}.
\newblock
  \url{https://www.bufferbloat.net/projects/bloat/wiki/BQL_enabled_drivers/},
  2014.

\bibitem{daps}
N.~Kuhn, E.~Lochin, A.~Mifdaoui, G.~Sarwar, O.~Mehani, and R.~Boreli.
\newblock Daps: Intelligent delay-aware packet scheduling for multipath
  transport.
\newblock In {\em IEEE International Conference on Communications (ICC)}, 2014.

\bibitem{dashDataset}
S.~Lederer, C.~M\"{u}ller, and C.~Timmerer.
\newblock Dynamic adaptive streaming over http dataset.
\newblock In {\em Proceedings of the 3rd Multimedia Systems Conference}, MMSys
  '12, 2012.

\bibitem{ecf}
Y.-s. Lim, E.~M. Nahum, D.~Towsley, and R.~J. Gibbens.
\newblock Ecf: An mptcp path scheduler to manage heterogeneous paths.
\newblock In {\em Proceedings of the 13th International Conference of CoNEXT},
  2017.

\bibitem{f2dpds}
D.~Ni, K.~Xue, P.~Hong, and S.~Shen.
\newblock Fine-grained forward prediction based dynamic packet scheduling
  mechanism for multipath tcp in lossy networks.
\newblock In {\em 2014 23rd International Conference on Computer Communication
  and Networks (ICCCN)}, 2014.

\bibitem{ocps}
D.~Ni, K.~Xue, P.~Hong, H.~Zhang, and H.~Lu.
\newblock Ocps: Offset compensation based packet scheduling mechanism for
  multipath tcp.
\newblock In {\em 2015 IEEE International Conference on Communications (ICC)},
  2015.

\bibitem{mptcpscheduler}
C.~Paasch, S.~Ferlin, O.~Alay, and O.~Bonaventure.
\newblock Experimental evaluation of multipath tcp schedulers.
\newblock In {\em Proceedings of the 2014 ACM SIGCOMM Workshop on Capacity
  Sharing Workshop}, CSWS '14, 2014.

\bibitem{mptcplinux}
C.~Paasch and B.~Sebastian.
\newblock {Multipath TCP in the Linux Kernel}.
\newblock \url{ http://www.multipath-tcp.org}, 2017.

\bibitem{queueawarecode}
{QAware}.
\newblock {QAware scheduler for MPTCPv0.93}.
\newblock \url{https://github.com/nitinder-mohan/mptcp-QueueAware.git}, 2018.

\bibitem{datacentermptcp}
C.~Raiciu, S.~Barre, C.~Pluntke, A.~Greenhalgh, D.~Wischik, and M.~Handley.
\newblock Improving datacenter performance and robustness with multipath tcp.
\newblock In {\em Proceedings of the ACM SIGCOMM 2011 Conference}.

\bibitem{opportunisticmptcp}
C.~Raiciu, D.~Niculescu, M.~Bagnulo, and M.~J. Handley.
\newblock Opportunistic mobility with multipath tcp.
\newblock In {\em Proceedings of the Sixth International Workshop on MobiArch},
  MobiArch '11, 2011.

\bibitem{rosberg}
Z.~Rosberg and P.~Kermani.
\newblock Customer routing to different servers with complete information.
\newblock {\em Advances in Applied Probability}, 21, 1989.

\bibitem{crosslayer_infocomm}
Y.~s.~Lim, Y.~C. Chen, E.~M. Nahum, D.~Towsley, and K.~W. Lee.
\newblock Cross-layer path management in multi-path transport protocol for
  mobile devices.
\newblock In {\em IEEE INFOCOM 2014 - IEEE Conference on Computer
  Communications}, 2014.

\bibitem{weber}
R.~R. Weber.
\newblock On the optimal assignment of customers to parallel servers.
\newblock {\em Journal of Applied Probability}, 1978.

\bibitem{whitt}
W.~Whitt.
\newblock Deciding which queue to join: Some counterexamples.
\newblock {\em Oper. Res.}, 1986.

\bibitem{winston}
W.~Winston.
\newblock Optimality of the shortest line discipline.
\newblock {\em Journal of Applied Probability}, 1977.

\bibitem{otias}
F.~Yang, Q.~Wang, and P.~D. Amer.
\newblock Out-of-order transmission for in-order arrival scheduling for
  multipath tcp.
\newblock In {\em 2014 28th International Conference on Advanced Information
  Networking and Applications Workshops}, 2014.

\end{thebibliography}

\end{document}